\documentclass[twocolumn,preprintnumbers,amsmath,amssymb,pra]{revtex4}

\usepackage{graphicx,amsmath}
\usepackage{dcolumn}
\usepackage{bm}
\usepackage{amssymb}
\usepackage{epstopdf}
\usepackage{color}

\begin{document}

\title{Spontaneous decay of artificial atoms in a three-qubit system}

\begin{abstract}
We study the evolution of qubits amplitudes in a one-dimensional
chain consisting of three equidistantly spaced noninteracting
qubits embedded in an open waveguide. The study is performed in
the frame of single-excitation subspace, where the only qubit in
the chain is initially excited. We show that the dynamics of
qubits amplitudes crucially depend on the value of $kd$, where $k$
is the wave vector, $d$ is a distance between neighbor qubits. If
$kd$ is equal to an integer multiple of $\pi$, then the qubits are
excited to a stationary level. In this case, it is the dark states
which prevent qubits from decaying to zero even though they do not
contribute to the output spectrum of photon emission. For other
values of $kd$ the excitations of qubits exhibit the damping
oscillations which represent the vacuum Rabi oscillations in a
three-qubit system. In this case, the output spectrum of photon
radiation is determined by a subradiant state which has the lowest
decay rate. We also investigated the case with the frequency of a
central qubit being different from that of the edge qubits. In
this case, the qibits decay rates can be controlled by the
frequency detuning between the central and the edge qubits.
\end{abstract}


\date{\today }

\author{Ya. S. Greenberg}\email{yakovgreenberg@yahoo.com}
\affiliation{Novosibirsk State Technical University, Novosibirsk,
Russia}
\author{A. A. Shtygashev} \affiliation{Novosibirsk State
Technical University, Novosibirsk, Russia}
\author{A. G. Moiseev} \affiliation{Novosibirsk State
Technical University, Novosibirsk, Russia}


 \maketitle

\section{Introduction}\label{intr}
Superconducting qubits coupled to photons propagating in an open
waveguide \cite{Asta2010,Hoi2011,Loo2013} allow for  the
investigation of the fascinating world of quantum light-matter
interactions in one dimension\cite{Roy2017,Gu2017,Shev2019}. Even
though the properties of multi-qubit 1D systems have been
extensively studied both theoretically
\cite{Alb2019,Zhang2019,Ruos2017,Lalum2013,Chang2012} and
experimentally \cite{Mirho2019,Brehm2021,Loo2014}, less attention
has been paid to a detail investigation of dynamic properties of
the few-qubit systems which are the building blocks for quantum
gates \cite{Barn2017,Kenfa2018}. Moreover, the scaling laws for
decay rates that have been found for multi-qubit systems
\cite{Tsoi2008} cannot obviously be applied for systems containing
few qubits.

As is known, the superconducting qubits can be technologically
addressed and controlled individually \cite{Shev2019}. Therefore,
it is important to know the evolution of the probability amplitude
of any qubit in a superconducting circuitry.

Here we explore the dynamic properties of a quantum circuit
consisting of a three-qubit linear chain, which is strongly
coupled to a common waveguide. The motivation for this choice is
that it is a simplest system with non-trivial properties for which
a full analytical treatment can be obtained. We investigate a
dynamic behavior of the qubits amplitudes with the only qubit in
the chain being initially excited. For three-qubit system, we find
the analytic expressions for the  complex energies and for the
collective states which define a temporal behavior of qubits
amplitudes. We show that the dynamics of qubits amplitudes
crucially depend on the value of $kd$, where $k$ is the wave
vector, $d$ is a distance between neighbor qubits. If $kd$ is
equal to integer multiple of $\pi$, the qubits are excited to a
stationary level. In this case, these are the dark states which
prevent qubits from decaying to zero even though they do not
contribute to the output photon spectrum. For other values of
$kd$, the excitations of qubits have oscillatory behavior and are
gradually damped out to zero. We also investigate the case with
the frequency of a central qubit being different from that of the
edge qubits. In this case, the qibits decay rates can be
controlled by the frequency detuning between the central and the
edge qubits. As the frequency detuning between central and edge
qubits increases, the decay rates of qubits are also increases.
This property is very important for the implementation of the
efficient control and readout protocols where a fast reset of the
excited qubits to their ground state is essential \cite{Zhou2021,
Maq2018}.

The paper is structured as follows.

In Sec. II, we begin by introducing a Jaynes-Cummings Hamiltonian
for atom-light interactions. We truncate the Hilbert space to a
single-excitation subspace and obtain a set of linear
integro-differential equations for the qubits amplitudes.

In Sec. III, in the frame of Wigner-Weisskopf approximation we
derive a set of linear differential equations for the qubits
amplitudes $\beta_n(t)$, which allow for a direct numerical
simulations.

A comprehensive analysis of the dynamics of the qubits amplitudes
is given in Sec. IV. We show that for $kd=n\pi$ the qubits
amplitudes become "frozen" at the constant level. The reason for
this is the dark states which prevent qubits from decaying to
zero. For the values of $kd$ which are not integer multiple of
$\pi$ the qubits amplitudes gradually damp out to zero. In this
section we also calculated the probability amplitude of the photon
emission. We find, that for $kd=(2n+1)\pi$ the evolution of the
photon amplitude consists of clearly seen steps. These steps can
be attributed to the interrelation between temporal behaviors of
the different qubits amplitudes.

In Sec. V we calculate a spectral density of photon radiation from
 a three-qubit chain. We show that for $kd=n\pi$ the dark states do not
 contribute to output radiation near resonance. In this case, the output
 spectral density has a Lorentzian lineshape. For $kd=(2n+1)\pi/2$ a spectral
 density exhibits two peaks which are a signature of the vacuum Rabi oscillations
 of qubits amplitudes. In general, if $kd$ is not equal to an integer
multiple of $\pi$, the width of a spectral line is defined by the
deep subradiant states.

In Sec. VI we formulate the dynamics of a three-qubit chain with
the aid of a non-Hermitian Hamiltonian which is obtained after the
elimination of the photon variables. We find the collective states
which are eigenvectors of non-Hermitian Hamiltonian and show how
the qubits amplitudes $\beta_n(t)$ can be expressed in terms of
these collective states.

In Sec. VII we consider a three-qubit system in which the
frequency $\Omega_0$ of the central qubit is different from that
of the edge qubits. We show that using the detuning
$\delta\Omega=\Omega-\Omega_0$ as external parameter we can
control the decay rates of the qubits' amplitudes. As the
frequency detuning between central and edge qubits increases, the
decay rates of qubits are also increases. For this case, a
spectral density of photon radiation is also calculated for
different  values of $\delta\Omega$.

 The main results of the paper are summarized in the concluding Section VIII.

\section{Formulation of the problem}

We consider a linear chain of three equally spaced qubits which are coupled to
photon field in an open waveguide (see Fig. 1).
\begin{figure}
  \includegraphics[width=6 cm]{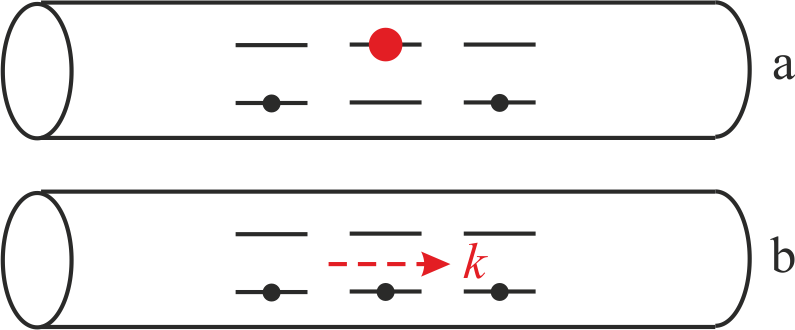}\\
  \caption{Schematic illustration of a single-excitation subspace for
  a three-qubit chain in an open waveguide. (a) A single qubit is excited, two
  qubits are in the ground state. (b) three qubits are in the ground state and
  a single photon propagates in the waveguide.}\label{FIG:Fig1}
\end{figure}
A distance between neighbor qubits is equal to $d$. The Hilbert
space of every qubit consists of the excited state $|e\rangle$ and
the ground state  $|g\rangle$. The Hamiltonian which accounts for
the interaction between qubits and the electromagnetic field is as
follows (we use $\hbar=1$ throughout the paper):

\begin{equation}\label{eq:math:1}
  H = H_0  + \sum\limits_k {\omega _k a_k^ +  a_k }+H_{int}
\end{equation}

where $H_0$ - is Hamiltonian of  bare qubits.

\begin{equation}\label{eq:math:a2}
H_0  = \frac{1}{2}\sum\limits_{n = 1}^3 {\left( {1 + \sigma
_z^{(n)} } \right)\Omega _n }
\end{equation}

\begin{equation}\label{eq:math:a1}
 H_{\operatorname{int} }
 = \sum\limits_{n = 1}^3 {} \sum\limits_k {} g_k^{(n)} e^{ - ikx_n }
 \sigma _ - ^{(n)} a_k^ +   + h.c.
\end{equation}

The quantity $g_k^{(n)}$  in \eqref{eq:math:a1} is the coupling
between $n$-th qubit and the photon field in a waveguide. Below we
consider a single-excitation subspace with either a single photon
is in a waveguide and all qubits are in the ground state,
Fig.~\ref{FIG:Fig1}b, or there are no photons in a waveguide with
the only n-th qubit in the chain being excited,
Fig.~\ref{FIG:Fig1}a. Therefore, we truncate  Hilbert space to the
following states:

\begin{equation}\label{eq:math:a3}
\begin{array}{l}
 \left| {G,1_k } \right\rangle  = \left| {g_1 ,g_2 ,g_3 } \right\rangle  \otimes \left| {1_k } \right\rangle ; \\
 \left| {1,0_k } \right\rangle  = \left| {e_1 ,g_2 ,g_3 } \right\rangle  \otimes \left| {0_k } \right\rangle ; \\
 \left| {2,0_k } \right\rangle  = \left| {g_1 ,e_2 ,g_3 } \right\rangle  \otimes \left| {0_k } \right\rangle ; \\
 \left| {3,0_k } \right\rangle  = \left| {g_1 ,g_2 ,e_3 } \right\rangle  \otimes \left| {0_k } \right\rangle  \\
 \end{array}
\end{equation}

The Hamiltonian \eqref{eq:math:a1} preserves the number of excitations
(number of excited qubits + number of photons).
 In our case the number of excitations is equal to one
 (see Fig.~\ref{FIG:Fig1}). Therefore, at any instant of time the system
 will remain
 within a single-excitation subspace.
 The wave function of an arbitrary single-excitation state can
 then be written in the form:

\begin{equation}\label{eq:math:a4}
\left| \Psi  \right\rangle  = \sum\limits_{n = 1}^3 {\beta _n
(t)e^{ - i\Omega _n t} } \left| {n,0_k } \right\rangle  +
\sum\limits_k {\gamma _k (t)e^{ - i\omega _k t} } \left| {G,1_k }
\right\rangle
\end{equation}
where $\beta_n(t)$  is the amplitude of $n$-th qubit, $\gamma_k(t)$  is
a single-photon amplitude which is related to a spectral density of spontaneous emission.

 \begin{equation}\label{eq:math:a5}
  S(\omega_k,t) =|\gamma_k(t)|^2.
\end{equation}

The function \eqref{eq:math:a4} is normalized to unity:

 \begin{equation}\label{eq:math:a6}
  \sum_{n=1}^3|\beta_n(t)|^2+\sum_k|\gamma_k(t)|^2 =1.
\end{equation}

From \eqref{eq:math:a6} we can find the full probability of photon
emission from the three-qubit system.

  \begin{equation}\label{eq:math:a7}
 P_{ph}(t)=\sum_k|\gamma_k(t)|^2=1- \sum_{n=1}^3|\beta_n(t)|^2.
\end{equation}
In fact, the quantity $P_{ph}(t)$ is the probability to find the
emitted photon at the moment $t$.

The equations for qubits amplitudes $\beta_n(t)$  in
\eqref{eq:math:a4}  can be found from time-dependent Schrodinger
equation  $id|\Psi\rangle/dt=H|\Psi\rangle$. For the amplitudes
$\gamma_k(t),\beta_n(t)$  we obtain:

\begin{equation}\label{eq:math:a8}
\begin{array}{l}
\frac{{d\beta _n }}{{dt}} =  - \sum\limits_k {|g_k^{(n)} |^2 }
 \int\limits_0^t {\beta _n (t')e^{ - i(\omega _k  - \Omega _n )(t - t')} dt'}  \\
- \sum\limits_{m \ne n}^3 {} \sum\limits_k {g_k^{*(n)} g_k^{(m)}
 e^{ - ik(x_m  - x_n )} e^{i(\Omega _n  - \Omega _m )t} }\\
 \times\int\limits_0^t {\beta _m (t')e^{ - i(\omega _k  - \Omega _m )(t - t')} dt'}  \\
 \end{array}
\end{equation}

\begin{equation}\label{eq:math:a9}
\gamma _k (t) =  - i\sum\limits_{n = 1}^3 {} g_k^{(n)} e^{ - ikx_n
} \int\limits_0^t {\beta _n (t')e^{i(\omega _k  - \Omega _n )t'}
dt'}
\end{equation}

According to \eqref{eq:math:a9} there are no photons in the system
at $t=0$. Our goal is to find the evolution of the amplitudes
$\beta_n(t)$   for any qubit in the chain when the only $n_0$-th
qubit ($n_0=1,2$, or $3$) is initially excited:

\begin{equation}\label{eq:math:a10}
\begin{cases}
&\beta_{n_0}(0)=1,\\
&\beta_{n}(0)=0,\;\;n\neq n_0.
\end{cases}
\end{equation}

\section{Equations for the qubits amplitudes}

We assume that the first and the third qubits are identical
($\Omega_1=\Omega_3\equiv\Omega$, $g_k^{(1)}=g_k^{(3)}\equiv
g_k$). The frequency and the coupling of the second qubit are
different ($\Omega_2\equiv\Omega_0$, $g_k^{(2)}\equiv g_k^{(0)})$.
A distance between central qubit and the edge qubits is equal to
$d$. We take the origin in the location of the second qubit:
$x_1=-d,\; x_2=0,\; x_3=+d$. In the frame of Wigner-Weisskopf
approximation, the equations \eqref{eq:math:a8} for the qubits
amplitudes can be reduced to the following set of linear
differential equations (see Appendix A for the derivation).

\begin{equation}\label{eq:math:a11}
\begin{array}{l}
\frac{{d\bar \beta _1 }}{{dt}} =  - \frac{\Gamma }{2}\bar \beta _1 (t) -
i\frac{{\Omega  - \Omega _0 }}{2}\bar \beta _1 (t) \\\\
- \bar \beta _2 (t)\frac{1}{2}\left( {\frac{{\Omega _0 }}{\Omega }}
\right)^{1/2} \sqrt {\Gamma \Gamma _0 } e^{ik_0 d}
 - \frac{\Gamma }{2}\bar \beta _3 (t)e^{2ikd}  \\
 \end{array}
\end{equation}

\begin{equation}\label{eq:math:a12}
\begin{array}{l}
 \frac{{d\bar \beta _2 }}{{dt}} =  - \frac{{\Gamma _0 }}{2}\bar \beta _2 (t)
  + i\frac{{\Omega  - \Omega _0 }}{2}\bar \beta _2 (t) \\
  \\
  - \frac{1}{2}\left( {\frac{\Omega }{{\Omega _0 }}}
  \right)^{1/2} \sqrt {\Gamma \Gamma _0 } e^{ikd} \left( {\bar \beta _1 (t)
   + \bar \beta _3 (t)} \right) \\
\end{array}
\end{equation}

 \begin{equation}\label{eq:math:a13}
\begin{array}{l}
 \frac{{d\bar \beta _3 }}{{dt}} =  - \frac{\Gamma }{2}\bar \beta _3 (t) - i\frac{{\Omega  - \Omega _0 }}{2}\bar \beta _3 (t) - \bar \beta _1 (t)\frac{\Gamma }{2}e^{2ikd}  \\
  \\
  - \bar \beta _2 (t)\frac{1}{2}\left( {\frac{{\Omega _0 }}{\Omega }} \right)^{1/2} \sqrt {\Gamma \Gamma _0 } e^{ik_0 d}  \\
 \end{array}
\end{equation}
where
\begin{equation}\label{eq:math:a14}
\begin{cases}
\bar{\beta}_{1,3}(t)&=e^{-i(\Omega-\Omega_0)t/2}{\beta}_{1,3}(t),\\
\bar{\beta}_{2}(t)&=e^{i(\Omega-\Omega_0)t/2}{\beta}_{2}(t).
\end{cases}
\end{equation}
$k=\Omega/v_g$, $k_0=\Omega_0/v_g$, $\Gamma$, $\Gamma_0$ are the
rates of spontaneous emission into the waveguide mode from edge
qubits and from the central qubit, respectively.

\section{The dynamics of three identical qubits}
For three identical qubits we obtain from
\eqref{eq:math:a11}-\eqref{eq:math:a13} the following equations:

\begin{equation}\label{eq:math:a15}
\frac{{d\beta _n }}{{dt}} =  - \frac{\Gamma }{2}\sum\limits_{m =
1}^3 {} \beta _m (t)e^{ikd\left| {m - n} \right|} ,\quad (n =
1,2,3)
\end{equation}
which can be expanded explicitly as:
\begin{equation}\label{15b}
\begin{array}{l}
 \frac{{d\beta _1 }}{{dt}} =  - \frac{\Gamma }{2}\beta _1 (t) - \frac{\Gamma }{2}\beta _2 (t)e^{ikd}  - \frac{\Gamma }{2}\beta _3 (t)e^{2ikd}
 \\\\
 \frac{{d\beta _2 }}{{dt}} =  - \frac{\Gamma }{2}\beta _2 (t) - \frac{\Gamma }{2}e^{ikd} \left( {\beta _1 (t) + \beta _3 (t)} \right)
 \\\\
 \frac{{d\beta _3 }}{{dt}} =  - \frac{\Gamma }{2}\beta _3 (t) - \frac{\Gamma }{2}\beta _1 (t)e^{2ikd}  - \frac{\Gamma }{2}\beta _2 (t)e^{ikd}  \\
 \end{array}
\end{equation}

where $k=\Omega/v_g$, and for simplicity we remove the bar over
$\beta_n(t)$.

The general solution of equations \eqref{15b} can be written as
follows:
\begin{equation}\label{16}
{\rm{\beta }}_i (t) = \sum\limits_{m = 1}^3 {} a_m^{(i)}
e^{\lambda _m t} ;\;i = 1,2,3
\end{equation}

The quantities $\lambda_m$ are characteristic roots, which can be
found by equating to zero the determinant of equations \eqref{15b}
\begin{equation}\label{17}
\left( {\lambda _i  + \frac{\Gamma }{2}} \right)\delta _{mn}  +
\frac{\Gamma }{2}e^{ikd\left| {m - n} \right|} \left( {1 - \delta
_{mn} } \right)
\end{equation}

The quantities $a_m^{(n)}$  in \eqref{16} are defined by initial
conditions:

\begin{equation}\label{eq:math:a17}
\begin{cases}
&\sum_{m=1}^3a_m^{(n_0)}=1,\\
&\sum_{m=1}^3a_m^{(i)}=0,\;\;i\neq n_0.
\end{cases}
\end{equation}
where $n_0$ is the number of initially excited qubit.

\begin{figure}[h]
  \includegraphics[width=8.5 cm]{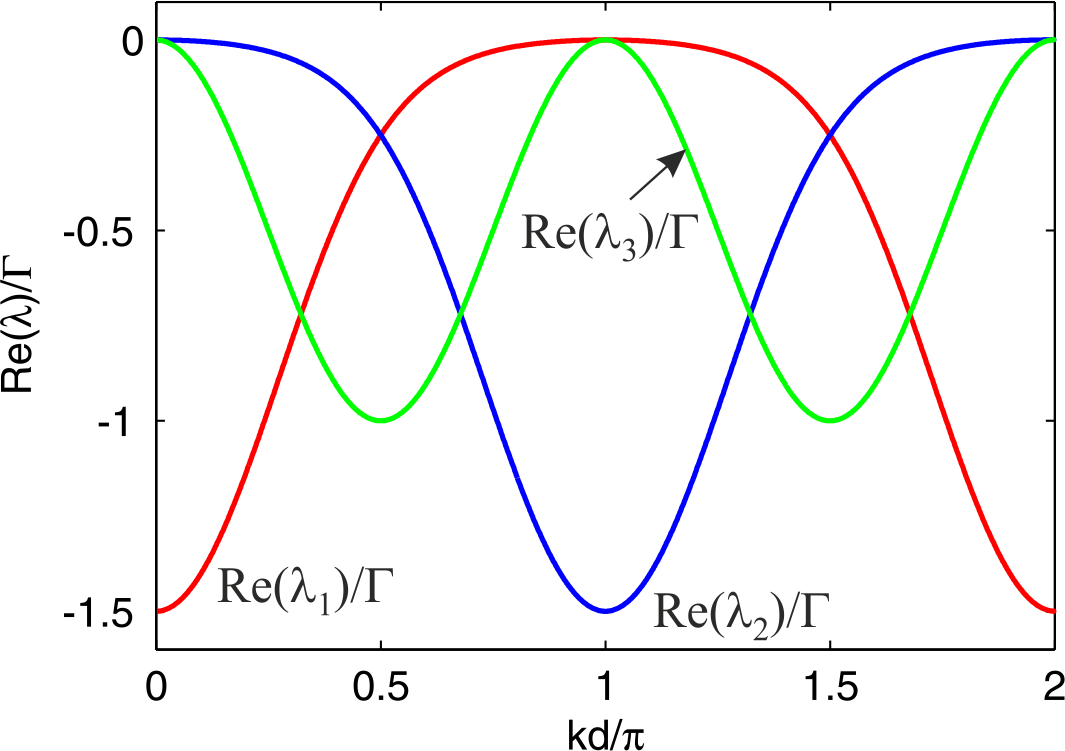}\\
  \caption{Dependence of $\mathrm{Re}(\lambda_i)$ on $kd$. $\mathrm{Re}(\lambda_1)$-red line, $\mathrm{Re}(\lambda_2)$-blue line, $\mathrm{Re}(\lambda_3)$-green line. All values of $\mathrm{Re}(\lambda_i)$ are negative.}\label{FIG:Fig2}
\end{figure}

From determinant of \eqref{17} we obtain the exact expressions for
$\lambda_m$:

\begin{equation}\label{eq:math:a18}
\begin{array}{*{20}c}
   {\lambda _1  =  - \frac{\Gamma }{4}e^{i2kd}  - \frac{\Gamma }{4}e^{ikd} \sqrt {e^{i2kd}  + 8}  - \frac{\Gamma }{2}}  \\
   \\
   {\lambda _2  =  - \frac{\Gamma }{4}e^{i2kd}  + \frac{\Gamma }{4}e^{ikd} \sqrt {e^{i2kd}  + 8}  - \frac{\Gamma }{2}}  \\
   \\
   {\lambda _3  = \frac{\Gamma }{2}e^{i2kd}  - \frac{\Gamma }{2}}  \\
\end{array}
\end{equation}
From \eqref{eq:math:a18} we see that the decay rates
$\Gamma_i=-\mathrm{Re}(\lambda_i)$ depend on $kd$. This dependence
is shown in Fig.~\ref{FIG:Fig2}.

In general, as is seen from \eqref{eq:math:a18}, all $\Gamma_i$'s
contribute to the decay rate of a concrete qubit. More important
is that a sum

\begin{equation}\label{eq:math:a19}
 \sum_{i=1}^3\lambda_i =-\frac{3}{2}\Gamma
\end{equation}
does not depend on $kd$. The expression \eqref{eq:math:a19}
is a special case of the more general sum rule for $N$ qubit system.

\begin{equation}\label{eq:math:a20}
 \sum_{i=1}^N\Gamma_i =N\Gamma
\end{equation}
where $\Gamma_i=2\mathrm{Re}(\lambda_i)$.
The sum rule \eqref{eq:math:a20} states that there are no other losses
in the system other than the coherent spontaneous emission into a waveguide.

Consider now the solution of equations \eqref{15b}. By subtracting
the third equation in \eqref{15b} from the first one we obtain:

\begin{equation}\label{eq:math:a21}
 \frac{d}{dt}(\beta_1-\beta_3) =-\frac{\Gamma}{2}\left(1-e^{i2kd}\right)(\beta_1-\beta_3)
\end{equation}
It follows from \eqref{eq:math:a21}  that if the edge qubits are initially not excited,  $\beta_1(0)-\beta_3(0)=0$, then this difference remains zero for all times,  $\beta_1(t)=\beta_3(t)\equiv\beta(t)$. This is quite reasonable from symmetry consideration: if a central qubit is initially excited, the temporal behavior of the amplitudes of the edge qubits must be the same.\\
Therefore, for this case, three equations \eqref{15b} can be
reduced to two equations:

\begin{equation}\label{eq:math:a22}
\begin{array}{l}
 \frac{{d\beta _2 }}{{dt}} =  - \frac{\Gamma }{2}\beta _2 (t) - \Gamma \beta (t)e^{ikd}  \\
\\
 \frac{{d\beta }}{{dt}} =  - \frac{\Gamma }{2}\beta (t)\left( {1 + e^{2ikd} } \right) - \frac{\Gamma }{2}\beta _2 (t)e^{ikd}  \\
 \end{array}
\end{equation}
where   $\beta_2(0)=1,\;\;\beta(0)=0$.\\

The characteristic roots of \eqref{eq:math:a22} are equal to
$\lambda_1$ and $\lambda_2$ which are given in
\eqref{eq:math:a18}. Therefore, the solution of equations
\eqref{eq:math:a22} reads:
 \begin{equation}\label{eq:math:a23}
\begin{cases}
\beta_2(t)&=b_1e^{\lambda_1t}+b_2e^{\lambda_2t},\\
\beta (t)&=a_1e^{\lambda_1t}+a_2e^{\lambda_2t},
\end{cases}
\end{equation}
where from initial conditions

\begin{equation}\label{eq:math:a24}
\begin{cases}
&b_1+b_2=1,\\
&a_1+a_2=0.
\end{cases}
\end{equation}

Another two conditions follow from \eqref{eq:math:a22} for time derivatives at $t=0$.

 \begin{equation}\label{eq:math:a25}
\begin{cases}
&b_1\lambda_1+b_2\lambda_2=-\cfrac{\Gamma}{2},\\
&a_1\lambda_1+a_2\lambda_2=-\cfrac{\Gamma}{2}e^{ikd}.
\end{cases}
\end{equation}

From \eqref{eq:math:a24} and \eqref{eq:math:a25} we obtain

 \begin{equation}\label{eq:math:a26}
b_1  =  - \frac{{\frac{\Gamma }{2} + \lambda _2 }}{{\lambda _1  -
\lambda _2 }};\;b_2  = \frac{{\frac{\Gamma }{2} + \lambda _1
}}{{\lambda _1  - \lambda _2 }}
\end{equation}

\begin{equation}\label{eq:math:a27}
a_1  =  - \frac{\Gamma }{2}\frac{{e^{ikd} }}{{\lambda _1  -
\lambda _2 }};\;a_2  = \frac{\Gamma }{2}\frac{{e^{ikd} }}{{\lambda
_1  - \lambda _2 }}
\end{equation}
Using the explicit expressions \eqref{eq:math:a18} we obtain

\begin{equation}\label{eq:math:a28}
\begin{array}{l}
 {{a}}_{\rm{1}} {\rm{ = }}\frac{1}{{{R}}}{{;a}}_{\rm{2}}  =  - \frac{1}{{R}} \\
 \\
 {{b}}_{\rm{1}} {\rm{ = }}\frac{{{{R}} - {\rm{e}}^{{\rm{ikd}}} }}{{{{2R}}}}{{;b}}_{\rm{2}}
 {\rm{ = }}\frac{{{{R + e}}^{{\rm{ikd}}} }}{{{{2R}}}} \\
 \end{array}
\end{equation}
where  $R=\sqrt{e^{i2kd}+8}$.

Now we assume that the first qubit in the chain is initially
excited.

\begin{equation}\label{eq:math:a29}
 \beta_1(0) =1,\;\beta_2(0)=0,\;\beta_3(0)=0.
\end{equation}
In this case, all amplitudes behave differently, so that the
equations \eqref{15b} should be used from which the solution can
be straightforwardly obtained

\begin{equation}\label{eq:math:a30}
\begin{array}{l}
 {\rm{\beta }}_{\rm{1}} {\rm{(t) = }}\frac{{{{b}}_{\rm{2}} }}{{\rm{2}}}e^{\lambda _1 t}  + \frac{{b_1 }}{2}e^{\lambda _2 t}  + \frac{1}{2}e^{_{\lambda _3 t} }  \\
 \\
 {\rm{\beta }}_{\rm{2}} {\rm{(t) = }}\frac{1}{{{R}}}\left( {{\rm{e}}^{{\rm{\lambda }}_{\rm{1}} {\rm{t}}}  - {\rm{e}}^{{\rm{\lambda }}_{\rm{2}} {\rm{t}}} } \right) \\
 \\
 {\rm{\beta }}_{\rm{3}} {\rm{(t) = }}\frac{{{{b}}_{\rm{2}} }}{{\rm{2}}}e^{\lambda _1 t}  + \frac{{b_1 }}{2}e^{\lambda _2 t}  - \frac{1}{2}e^{_{\lambda _3 t} }  \\
 \end{array}
\end{equation}
where $b_1$ and $b_2$ are given in \eqref{eq:math:a28}. For
special cases when $kd$ is integer multiple of $\pi$ we may obtain
from \eqref{eq:math:a23} and \eqref{eq:math:a30} very simple forms
for the temporal behavior of the qubits' amplitudes.

\subsection{$kd=\pi n$}

For this case, $\lambda_1=-\frac{3}{2}\Gamma$,
$\lambda_2=\lambda_3=0$ if $n$ is even number, and $\lambda_1=0$,
$\lambda_2=-\frac{3}{2}\Gamma$, $\lambda_3=0$ if $n$ is odd
number. The calculations show that no matter which qibit in the
chain is initially excited (central or edge qubit) its evolution
is the same.

\begin{equation}\label{eq:math:a31}
\beta_{exc}(t)=\frac{1}{3}e^{-3\Gamma t/2}+\frac{2}{3}
\end{equation}

The evolution of unexcited qubits are also the same within a phase factor of $\pi$.

\begin{equation}\label{eq:math:a32}
\beta_{unexc}(t)=(-1)^n\left(\frac{1}{3}e^{-3\Gamma t/2}-\frac{1}{3}\right)
\end{equation}

The expression \eqref{eq:math:a32} is valid for any $n$ if the
central qubit is excited. If the first qubit is excited and $n$ is
the odd number, the amplitudes of unexcited qubits evolve with the
opposite phases:  $\beta_2(t)=\beta_{unexc}(t)$,
$\beta_3(t)=-\beta_{unexc}(t)$.

It worth noting the equality of the amplitudes of the second and
the third qubit (if the first qubit is initially excited) even
though they are located at different distance from the excited
qubit. This is the consequence of Wigner-Weisskopf (or Markov)
approximation: in this special case there are no interference
between qubits so that all unexcited qubits feel the photon field
simultaneously no matter how far they are from the excited qubit.
From \eqref{eq:math:a31} and \eqref{eq:math:a32} we find a full
probability of photon emission \eqref{eq:math:a7}

 \begin{equation}\label{eq:math:a33}
P_{ph}(t)=1-2|\beta_{unexc}(t)|^2-|\beta_{exc}(t)|^2=\frac{1}{3}\left(1-e^{-3\Gamma t}\right)
\end{equation}

\begin{figure}[h]
  \includegraphics[width=8 cm]{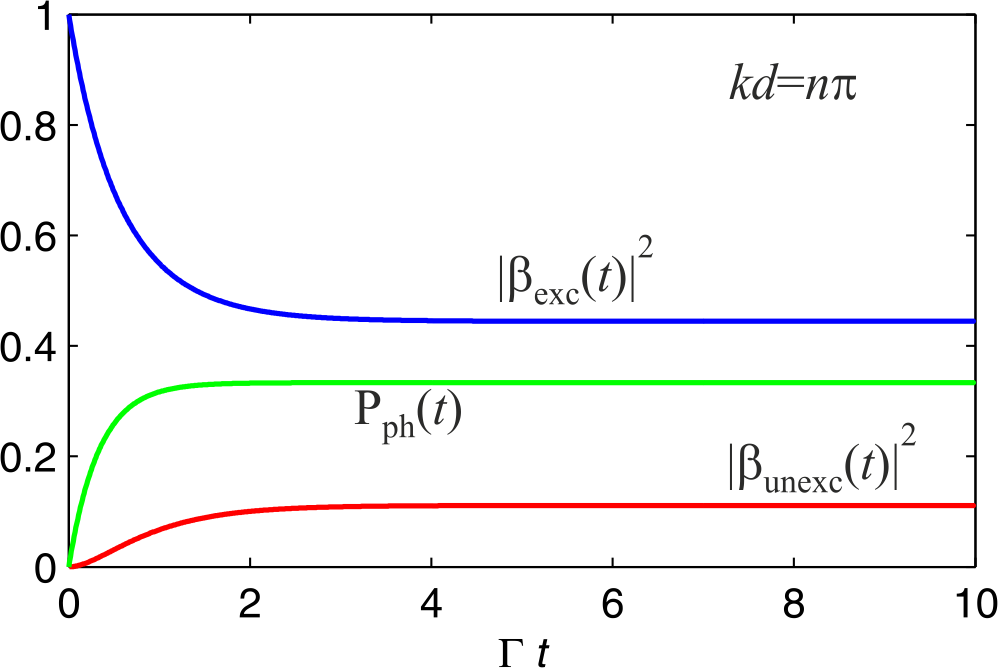}\\
  \caption{The probability amplitudes for $kd=n\pi$ of excited,
$|\beta_{exc}(t)|^2$ (the upper line) and unexcited
$|\beta_{unexc}(t)|^2$  (the lower line) qubits. The probability
of a full photon emission  is shown between the qubit
lines.}\label{FIG:Fig3}
\end{figure}

The probability of qubits amplitudes and a full probability of
photon emission are shown in Fig.~\ref{FIG:Fig3}

As is seen from Fig.~\ref{FIG:Fig3}, as the time proceeds the
qubits amplitudes become "frozen" at some level. We see from
\eqref{eq:math:a31} and \eqref{eq:math:a32} that this frozenness
is solely due to the dark states ($\lambda_n=0$), which prevent
the qubits' amplitudes from decaying to zero. The frozenness can
be lifted if $kd$ is not equal to an integer multiple of $\pi$.
Then all qubits are damped to zero with the rate being determined
by the root of $\lambda_n$ which has the lowest real part.

\subsection{$kd=\pi (2n+1)/2$}

If n is even number then
$\lambda_1=-\frac{\Gamma}{4}(1+i\sqrt{7})$,
$\lambda_2=-\frac{\Gamma}{4}(1-i\sqrt{7})$, $\lambda_3=-\Gamma$ ,
and if n is odd number then
$\lambda_1=-\frac{\Gamma}{4}(1-i\sqrt{7})$,
$\lambda_2=-\frac{\Gamma}{4}(1+i\sqrt{7})$, $\lambda_3=-\Gamma$

If the central qubit is initially excited we obtain:

\begin{equation}\label{eq:math:a34}
\begin{array}{l}
 \beta _1 (t) = \beta _3 (t) = ( - 1)^{n + 1} i\frac{2}{{\sqrt 7 }}e^{ - \frac{1}{4}\Gamma t} \sin \left( {\frac{{\sqrt 7 }}{4}\Gamma t} \right) \\
 \\
 \beta _2 (t) = e^{ - \frac{1}{4}\Gamma t} \left( {\cos \left( {\frac{{\sqrt 7 }}{4}\Gamma t} \right) + ( - 1)^n \frac{1}{{\sqrt 7 }}\sin \left( {\frac{{\sqrt 7 }}{4}\Gamma t} \right)} \right) \\
 \end{array}
\end{equation}

For initially excited edge qubit we obtain:

\begin{equation}\label{eq:math:a35}
\begin{array}{l}
 \beta _1 (t) = \frac{{e^{ - \frac{1}{4}\Gamma t} }}{2}\left( {\cos \left( {\frac{{\sqrt 7 }}{4}\Gamma t} \right) + \frac{1}{{\sqrt 7 }}\sin \left( {\frac{{\sqrt 7 }}{4}\Gamma t} \right)} \right) + \frac{1}{2}e^{ - \Gamma t}  \\
\\
 \beta _2 (t) = ( - 1)^{n + 1} i\frac{2}{{\sqrt 7 }}e^{ - \frac{1}{4}\Gamma t} \sin \left( {\frac{{\sqrt 7 }}{4}\Gamma t} \right) \\
\\
 \beta _3 (t) = \frac{{e^{ - \frac{1}{4}\Gamma t} }}{2}\left( {\cos \left( {\frac{{\sqrt 7 }}{4}\Gamma t} \right) + \frac{1}{{\sqrt 7 }}\sin \left( {\frac{{\sqrt 7 }}{4}\Gamma t} \right)} \right) - \frac{1}{2}e^{ - \Gamma t}  \\
 \end{array}
\end{equation}

The oscillatory behavior of the qubits amplitudes in these
expressions is a signature of the vacuum Rabi oscillations with
the frequency  $\sqrt{7}\Gamma/4$.

\begin{figure}[h]
  \includegraphics[width=8 cm]{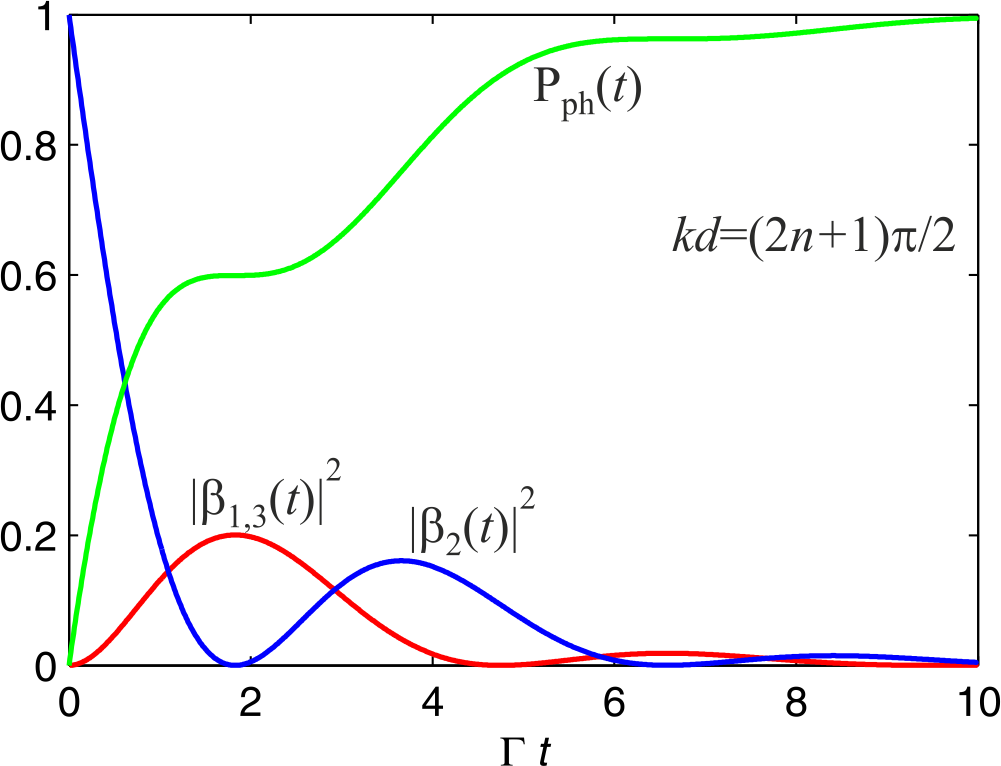}\\
  \caption{The evolution of qubits amplitudes for $kd$ equals
  the half-integer multiple of $\pi$ with the central qubit
  being initially excited.}\label{Fig4}
\end{figure}

The evolution of qubits amplitudes for half-integer multiple of
$\pi$ with the central qubit being initially excited is shown in
Fig.~\ref{Fig4}.

\begin{figure}[h]
  \includegraphics[width=8 cm]{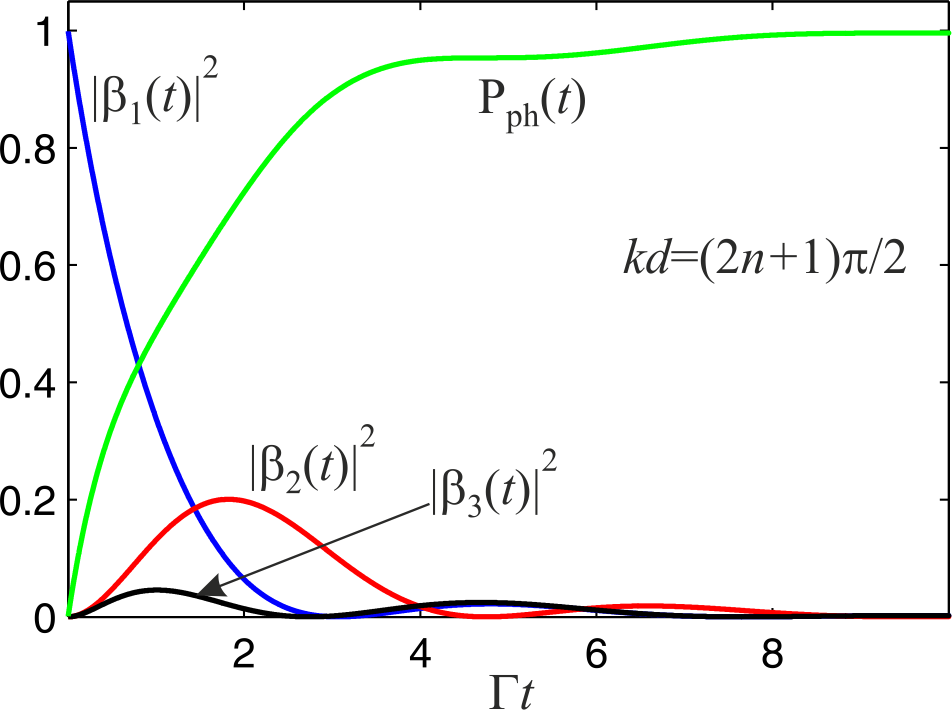}\\
  \caption{The evolution of qubits amplitudes for $kd$ equals the
  half-integer multiple of $\pi$ with the first qubit being
  initially excited}\label{FIG:Fig5}
\end{figure}

The evolution of $P_{ph}(t)$  reveals two clear visible steps
where,  $dP_{ph}(t)/dt=0$. These steps can be attributed to the
interrelation between the temporal dynamics of the different
amplitudes. As is seen from this figure, the steps on $P_{ph}(t)$
curve are in the vicinity of extremum points of qubits amplitudes.
Physically, these steps are the signature of the trapping of the
photon radiation. As is seen in Fig.~\ref{FIG:Fig3}, near the
first step the radiation emitted by excited qubit is absorbed by
unexcited qubits, so that the rate of the output radiation is not
changed. The evolution of qubits' amplitudes for half-integer
multiple of $\pi$ with the first qubit being initially excited is
shown in Fig.~\ref{FIG:Fig5}.

Here, we see that the evolution of qubits amplitude depends on
their distance from the excited qubit.  The greater is the
distance of a qubit from the excited one, the less it is affected
by the excitation.

\section{Spectral density of photon radiation}

The quantity $\gamma_k(t)$  in \eqref{eq:math:a4} allows
for the calculation of a spectral density \eqref{eq:math:a5} of
spontaneous emission into a waveguide, $S(\omega,t)$.
The equation \eqref{eq:math:a9} for identical qubits reads:

 \begin{equation}\label{eq:math:a36}
\gamma (\omega ,t) =  - ig_k^{} \sum\limits_{n = 1}^3 {} e^{ -
ikx_n } \int\limits_0^t {\beta _n (t')e^{i(\omega  - \Omega )t'}
dt'}
\end{equation}

Using the expression \eqref{16} for $\beta_n(t)$ we obtain

\begin{equation}\label{eq:math:a37}
\gamma (\omega ,t) = g_k^{} \sum\limits_{n,m = 1}^3 {} a_j^{(n)}
e^{ - ikx_n } \frac{{1 - e^{i(\omega  - \Omega  - i\lambda _m )t}
}}{{\omega  - \Omega  - i\lambda _m }}
\end{equation}

As $g_k$ in (\ref{eq:math:a36}) and (\ref{eq:math:a37}) depends on
the waveguide length, $L$ and the light velocity $v_g$ (see
(\ref{A15a})), we factorize $\gamma_k(t)$ as follows:

\[
\gamma _k (t) \equiv \left( {\frac{{v_g }} {{2L\Omega }}}
\right)^{1/2} \left( {\frac{\Gamma } {\Omega }} \right)^{1/2} f_k
(t)
\]

Therefore, we define spectral density in the following form:
\begin{equation}\label{Sd}
S(\omega ,t) = \frac{{\left| {\gamma _k (t)} \right|^2 }} {{\left(
{\frac{{v_g }} {{2L\Omega }}} \right)}} = \frac{\Gamma } {\Omega
}\left| {f_k (t)} \right|^2
\end{equation}

The equation (\ref{eq:math:a37}) is a rather general expression.
As the time proceeds only the term with the lowest real part of
$\lambda_m$ survives. The concise analytical results can be
obtained only for several simple cases. Below we calculate the
spectral density for $kd=n\pi$, using the amplitudes
\eqref{eq:math:a31} and \eqref{eq:math:a32} in
\eqref{eq:math:a36}.

When a central qubit is initially excited we obtain for
$\gamma_k(t)$ the following result:

\begin{equation}\label{eq:math:a38}
\begin{array}{l}
 \gamma _k (t) =  - g_k \frac{2}{3}\left( {1 - ( - 1)^n \cos \left( {\frac{\omega }{\Omega }n\pi } \right)} \right)\frac{{e^{i(\omega  - \Omega )t}  - 1}}{{\omega  - \Omega }} \\
 \\
  - g_k \frac{1}{3}\left( {1 + ( - 1)^n 2\cos \left( {\frac{\omega }{\Omega }n\pi } \right)} \right)\frac{{e^{i\left( {\omega  - \Omega  + i\frac{{3\Gamma }}{2}} \right)t}  - 1}}{{\left( {\omega  - \Omega  + i\frac{{3\Gamma }}{2}} \right)}} \\
 \end{array}
\end{equation}

If the first qubit is initially excited we obtain:

\begin{equation}\label{eq:math:a39}
\begin{array}{l}
 \gamma _k (t) =  - g_k \frac{1}{3}\left( {2e^{i\left( {\frac{\omega }{\Omega }n\pi } \right)}  - e^{ - i\left( {\frac{\omega }{\Omega }n\pi } \right)}  - ( - 1)^n } \right)\frac{{e^{i(\omega  - \Omega )t}  - 1}}{{\omega  - \Omega }} \\
 \\
  - g_k \frac{1}{3}\left( {( - 1)^n  + 2\cos \left( {\frac{\omega }{\Omega }n\pi } \right)} \right)\frac{{e^{i\left( {\omega  - \Omega  + i\frac{{3\Gamma }}{2}} \right)t}  - 1}}{{\left( {\omega  - \Omega  + i\frac{{3\Gamma }}{2}} \right)}} \\
 \end{array}
\end{equation}

It worth noting that in expressions
\eqref{eq:math:a38},\eqref{eq:math:a39} a wave vector $k$ is
related to a running frequency $\omega$:  $k=\omega/v_g$. The
first lines in \eqref{eq:math:a38} and \eqref{eq:math:a39} are
related to the dark states ($\lambda_n=0$). They do not lead to a
singularity near resonance  $\omega\approx\Omega$, since $kd$ -
dependent prefactors in this terms tend to zero more rapidly than
the denominator does. At the point of a resonance these terms
exactly equal to zero. A physical reason for this is that the dark
states do not interact with a photon field and, therefore, cannot
contribute to the photon emission. Therefore, near the resonance,
a spectral density of spontaneous emission can be approximated by
a Lorentzian form with a full width at half the height of the
resonance line being equal to $3\Gamma$.

\begin{equation}\label{eq:math:a40}
\gamma _k (t) \approx g_k \frac{{e^{i\left( {\omega  - \Omega  +
i\frac{{3\Gamma }} {2}} \right)t}  - 1}} {{\left( {\omega  -
\Omega  + i\frac{{3\Gamma }} {2}} \right)}}\xrightarrow[{t \to
\infty }]{}\frac{{g_k }} {{\left( {\omega  - \Omega  +
i\frac{{3\Gamma }} {2}} \right)}}
\end{equation}

\begin{equation}\label{eq:math:a41}
S(\omega ,t \to \infty ) =  \frac{\Gamma \Omega}{{(\omega  -
\Omega )^2 + \left( {\frac{{3\Gamma }}{2}} \right)^2 }}
\end{equation}

For $kd=(2n+1)\pi/2 $, the qubits amplitudes are given in
equations (\ref{eq:math:a34}) and (\ref{eq:math:a35}). The
calculation of (\ref{eq:math:a36}) with these amplitudes results
in the following expressions for $\gamma_k(t)$ with the central
qubit being initially excited:

\begin{equation}\label{39}
\begin{gathered}
  \gamma _k (t) =  - g_k \left[ {\frac{1}
{2}\left( {1 - \frac{{( - 1)^n i}} {{\sqrt 7 }}} \right) -
\frac{{2( - 1)^n }}
{{\sqrt 7 }}\cos kd} \right] \hfill \\
   \times \frac{{e^{i(\omega  - \Omega  + \frac{{\sqrt 7 }}
{4}\Gamma  + i\frac{1} {4}\Gamma )t}  - 1}} {{\omega  - \Omega  +
\frac{{\sqrt 7 }} {4}\Gamma  + i\frac{1}
{4}\Gamma }} \hfill \\
   - g_k \left[ {\frac{1}
{2}\left( {1 + \frac{{( - 1)^n i}} {{\sqrt 7 }}} \right) +
\frac{{2( - 1)^n }}
{{\sqrt 7 }}\cos kd} \right] \hfill \\
   \times \frac{{e^{i(\omega  - \Omega  - \frac{{\sqrt 7 }}
{4}\Gamma  + i\frac{1} {4}\Gamma )t}  - 1}} {{\omega  - \Omega  -
\frac{{\sqrt 7 }} {4}\Gamma  + i\frac{1}
{4}\Gamma }} \hfill \\
\end{gathered}
\end{equation}

If the edge qubit is initially excited we obtain:

\begin{equation}\label{40}
\begin{gathered}
  \gamma _k (t) =  - ig_k \left[ {\frac{1}
{2}\left( {1 - \frac{i} {{\sqrt 7 }}} \right)\cos kd + \frac{{( -
1)^n }}
{{\sqrt 7 }}} \right] \hfill \\
   \times \frac{{e^{i(\omega  - \Omega  + \frac{{\sqrt 7 }}
{4}\Gamma  + i\frac{1} {4}\Gamma )t}  - 1}} {{\omega  - \Omega  +
\frac{{\sqrt 7 }} {4}\Gamma  + i\frac{1}
{4}\Gamma }} \hfill \\
   - ig_k \left[ {\frac{1}
{2}\left( {1 + \frac{i} {{\sqrt 7 }}} \right)\cos kd - \frac{{( -
1)^n }}
{{\sqrt 7 }}} \right] \hfill \\
   \times \frac{{e^{i(\omega  - \Omega  - \frac{{\sqrt 7 }}
{4}\Gamma  + i\frac{1} {4}\Gamma )t}  - 1}} {{\omega  - \Omega  -
\frac{{\sqrt 7 }} {4}\Gamma  + i\frac{1}
{4}\Gamma }} \hfill \\
   - ig_k \sin kd\frac{{e^{i(\omega  - \Omega  + i\frac{1}
{2}\Gamma )t}  - 1}} {{\omega  - \Omega  + i\frac{1}
{2}\Gamma }} \hfill \\
\end{gathered}
\end{equation}

In expressions (\ref{39}), (\ref{40})
$kd=\frac{\omega}{\Omega}(2n+1)\pi/2$ .

\begin{figure}[h]
  \includegraphics[width=8 cm]{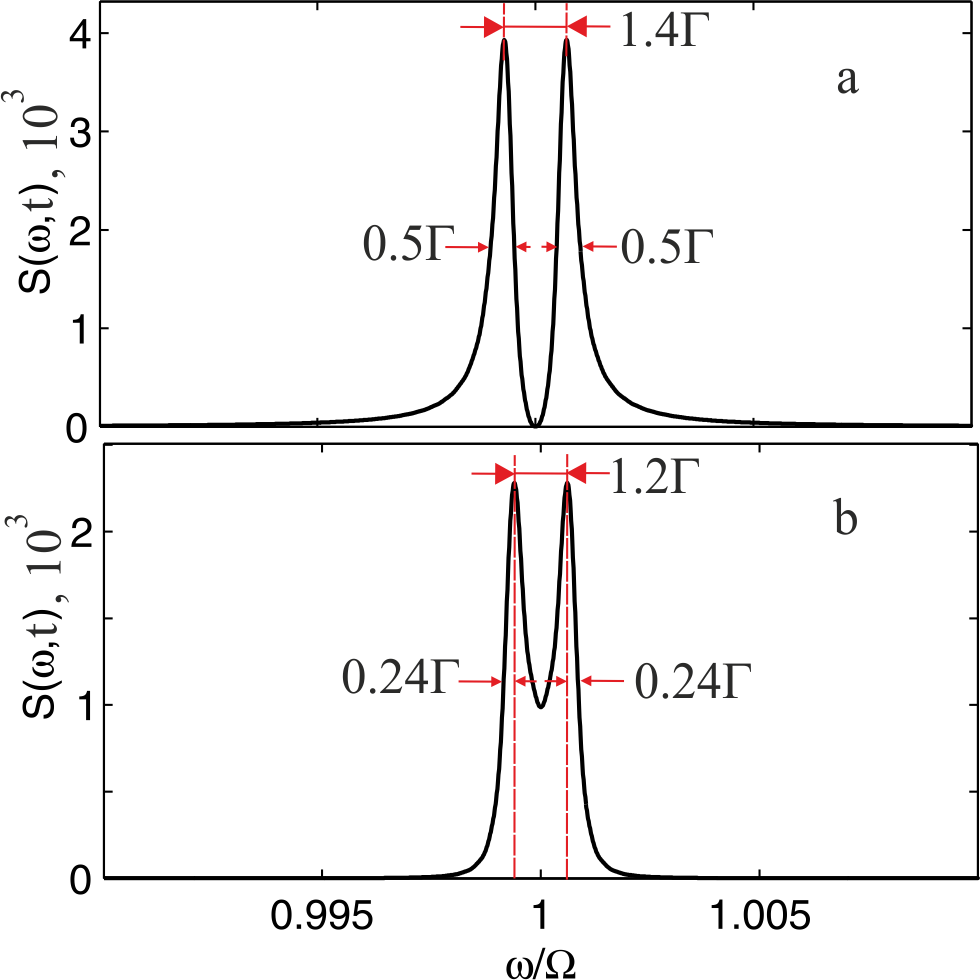}\\
  \caption{Spectral density for three-qubit system
  calculated from (\ref{39}) and (\ref{40}) for $n=1, kd=1.5\pi , t=20/\Gamma,
  \Gamma/\Omega=10^{-3}$. (a) second qubit is initially excited, (b) first qubit is initially
excited.}\label{FIG6}
\end{figure}

The spectral density of photon radiation (\ref{Sd}) calculated
from (\ref{39}) and (\ref{40}) for $n=1, kd=1.5\pi , t=20/\Gamma,
\Gamma/\Omega=10^{-3}$ is shon in Fig. \ref{FIG6}.
Two peaks at these plots are a clear signature of vacuum Rabi
oscillations..

\section{Effective non-Hermitian Hamiltonian and collective
states}

The elimination of photon variables allows us to express the
photon-mediated interaction between identical qubits in terms of a
non-Hermitian effective Hamiltonian, which in the Markovian
approximation reads \cite{Chang2012,Lehm1970}:

\begin{equation}\label{41}
H_{eff}  =  - i\frac{\Gamma } {2}\sum\limits_{m,n = 1}^3 {}
e^{ik\left| {x_m  - x_n } \right|} \sigma _m^ +  \sigma _n
\end{equation}

where $k= \Omega/v_g$, $v_g$ is the velocity of electromagnetic
wave in a waveguide, $x_n$ is the position of $n$-th qubit,
$\sigma^+_n, \sigma_n$ are raising and lowering spin operators for
$n$-th qubit.

The rate of spontaneous emission $\Gamma$ of an individual qubit
is defined by the Fermi golden rule:

\begin{equation}\label{42}
\Gamma  = 2\pi \sum\limits_k {\left| {g_k^{} } \right|^2 \delta
(\omega _k  - \Omega )}
\end{equation}

It follows from (\ref{41}) that the photon-mediated interaction
between qubits in such a system results in the coherent $ J_{mn} =
\Gamma \sin \left( {k\left| {x_m  - x_n } \right|} \right)/2$ and
dissipative $\Gamma _{mn}  = \Gamma \cos \left( {k\left| {x_m -
x_n } \right|} \right)$ rates. The coherent rate shifts the
positions of the qubits resonances, while the dissipative rate
gives rise to the additional spontaneous emission into the
waveguide mode. Unlike the real atoms with short-range
dipole-dipole interaction, here a coherent interaction $J_{mn}$ is
a long-range one: every qubit is sensitive to its distant
neighbor.

The wave function for Hamiltonian (\ref{41}) can be expressed in
terms of a superposition of the single excited states

\begin{equation}\label{43}
\Psi(t) = \sum\limits_{n = 1}^3\beta_n(t)\left|n\right\rangle
\end{equation}

Even though the wave functions (\ref{eq:math:a4}) and (\ref{43})
are different, the qubits amplitudes $\beta_n(t)$  in these
expressions are the same quantities`. Indeed, the equations for
$\beta_n(t)$ in (\ref{43}) can be derived from the time-dependent
Schrodinger equation ${{id\Psi /dt = H}}_{{{eff}}} {{\Psi }}$ :

\begin{equation}\label{44}
\frac{{d\beta _n }} {{dt}} =  - \frac{\Gamma } {2}\sum\limits_{m =
1}^3 {} \beta _m (t)e^{ik\left| {x_m  - x_n } \right|} ,\quad (n =
1,2,3)
\end{equation}

The equations (\ref{44}) are nothing but a set of equations
(\ref{15b}). Therefore, within a single excitation subspace two
Hamiltonians, (\ref{eq:math:1}) and (\ref{41}) are equivalent in
that they provide the same equations for the qubits amplitudes.

In a single-excitation subspace, the Hamiltonian (\ref{41}) has
three collective eigenfunctions:

\begin{equation}\label{45}
\left| {\Psi _i (t)} \right\rangle  = e^{ - i\bar E_i t}
\sum\limits_{n = 1}^3 {} \alpha _n^{(i)} \left| n \right\rangle
;\;(i = 1,2,3)
\end{equation}

where $\bar E_i$ is a complex energy
\begin{equation}\label{46}
\bar E_i  = E_i  - i\frac{{\Gamma _i }} {2}
\end{equation}

The quantities $E_i$ and $\Gamma_i$ depend on the system
parameters $\Gamma, k, x_n$. The wave vectors with
$\Gamma_i<\Gamma$ are called subradiant states, those with
$\Gamma_i>\Gamma$ are called superradiant states. The coefficients
in (\ref{45}) can be obtained from the Schrodinger equation $
H_{eff} \Psi  = E\Psi$.

\begin{equation}\label{47}
E_j \sum\limits_{n = 1}^3 {} \alpha _n^{(j)} \left| n
\right\rangle =  - i\frac{\Gamma } {2}\sum\limits_{m,n = 1}^3 {}
e^{ikd\left| {m - n} \right|} \alpha _n^{(j)} \left| m
\right\rangle ; j = 1,2,3
\end{equation}

Complex energies can be found by equating to zero the determinant
of the matrix:

\begin{equation}\label{48}
\left( {E + i\frac{\Gamma } {2}} \right)\delta _{mn}  +
i\frac{\Gamma } {2}e^{ikd\left| {m - n} \right|} \left( {1 -
\delta _{mn} } \right);\quad (m,n = 1,2,3)
\end{equation}

A comparison between the determinants of (\ref{17}) and (\ref{48})
shows that the quantities  $\lambda_n$ are related to those of
complex energies, $\bar E_n$ : $\bar E_n=i\lambda_n$ where
$\lambda_n$ are given in (\ref{eq:math:a18}).

From (\ref{47}) we obtain three equations for the coefficients,  .

\begin{equation}\label{50}
\begin{gathered}
  \left( {E_j  + i\frac{\Gamma }
{2}} \right)\alpha _1^{(j)}  + i\frac{\Gamma } {2}e^{ikd} \alpha
_2^{(j)}  + i\frac{\Gamma }
{2}e^{2ikd} \alpha _3^{(j)}  = 0 \hfill \\
  \left( {E_j  + i\frac{\Gamma }
{2}} \right)\alpha _2^{(j)}  + i\frac{\Gamma } {2}e^{ikd} \alpha
_1^{(j)}  + i\frac{\Gamma }
{2}e^{ikd} \alpha _3^{(j)}  = 0 \hfill \\
  \left( {E_j  + i\frac{\Gamma }
{2}} \right)\alpha _3^{(j)}  + i\frac{\Gamma } {2}e^{ikd} \alpha
_2^{(j)}  + i\frac{\Gamma }
{2}e^{2ikd} \alpha _1^{(j)}  = 0 \hfill \\
\end{gathered}
\end{equation}
where for simplicity we remove the bar over $E_i$.

Because Hamiltonian (\ref{41}) is non-Hermitian, the
eigenfunctions (\ref{45}) are neither normalized nor orthonormal.
It is known that a correct calculation of the coefficients
$\alpha_n^{(i)}$ in (\ref{45}) requires a bi-orthogonal set of
eigenfunctions $ \left| {\bar \Psi _i (t)} \right\rangle$ which
are a solution of the Schrodinger equation for $H_{eff}^\dag$. In
our case $H_{eff}^\dag = H_{eff}^ *$ with the consequence that the
complex conjugate of an eigenstate $\left|{\Psi_i (t)}
\right\rangle$ of $H_{eff}$ is an eigenstate of $H_{eff}^\dag$.
Therefore, the conditions for normalization and orthonormality
between eigenfunctions of these two sets lead to the following
equations for the coefficients \cite{Keck2003,Brody2014}.

\begin{equation}\label{51a}
\sum\limits_{n = 1}^3 {} \left( {\alpha _n^{(i)} } \right)^2  =
1,\quad i = 1,2,3
\end{equation}

\begin{equation}\label{51b}
\sum\limits_{n = 1}^3 {} \alpha _n^{(i)} \alpha _n^{(j)}  =
0,\quad i \ne j,\quad i,j = 1,2,3
\end{equation}

It worth noting that the coefficients $\alpha_n^{(i)}$  in
(\ref{51a}) and (\ref{51b}) are in general complex quantities. In
what follows we use equations (\ref{50}) and the conditions
(\ref{51a}) and (\ref{51b}) for the calculation of the
coefficients $\alpha_n^{(i)}$.

As the energies $ E_j$ are obtained from the determinant of
equations (\ref{50}), these equations are not independent.
Therefore, for the calculations of the coefficients
$\alpha_n^{(i)}$ we may take any two of them. For subsequent
calculations we take first and second equations in (\ref{50}).
First, we calculate the coefficients $\alpha_n^{(1)}$ . With
$E_1=i\lambda_1$ determined in (\ref{eq:math:a18}) two first
equations in (\ref{50}) read as follows:

\begin{equation}\label{52}
\begin{gathered}
  \left( {e^{ikd}  + R} \right)\alpha _1^{(1)}  - 2\alpha _2^{(1)}  - 2e^{ikd} \alpha _3^{(1)}  = 0 \hfill \\
  \left( {e^{ikd}  + R} \right)\alpha _2^{(1)}  - 2\alpha _1^{(1)}  - 2\alpha _3^{(1)}  = 0 \hfill \\
\end{gathered}
\end{equation}
where $R$ is given in (\ref{eq:math:a28}).

Additional third equation is given by the normalizing condition
(\ref{51a}). Therefore, for the wave-function $\left| {\Psi _1
(t)} \right\rangle$ we find the following set of the coefficients
$\alpha_n^{(1)}$ .

\begin{equation}\label{53a}
\alpha _1^{(1)}  = \alpha _3^{(1)}  =  \pm \frac{{e^{ikd} \left(
{e^{ikd}  + R} \right) + 2}} {D(R)}
\end{equation}

\begin{equation}\label{53b}
\alpha _2^{(1)}  =  \pm \frac{{3e^{ikd}  +R}} {D(R)}
\end{equation}

where
\begin{equation}\label{D}
D(R) = \left( {4e^{4ikd}  + 34e^{2ikd}  + 14e^{ikd} R + 4e^{3ikd}
R + 16} \right)^{1/2}
\end{equation}

Similar calculations for $E_2=i\lambda_2$ provide the result that
differs from (\ref{53a}), (\ref{53b}) only by the sign of $R$.
Therefore, for the wave-function  $\left| {\Psi _2 (t)}
\right\rangle$ we find the following set of the coefficients
$\alpha_n^{(2)}$ .

\begin{equation}\label{54a}
\alpha _1^{(2)}  = \alpha _3^{(2)}  =  \pm \frac{{e^{ikd} \left(
{e^{ikd}  - R} \right) + 2}} {D(-R)}
\end{equation}

\begin{equation}\label{54b}
\alpha _2^{(2)}  =  \pm \frac{{3e^{ikd}  - R}} {D(-R)}
\end{equation}

The signs in right hand side of equations (\ref{53a}),
(\ref{53b}), and (\ref{54a}), (\ref{54b}) must be the same (plus
or minus) for all four expressions. This requirement follows from
the orthonormality condition (\ref{51b}):

\begin{equation}\label{55}
\alpha _1^{(1)} \alpha _1^{(2)}  + \alpha _2^{(1)} \alpha _2^{(2)}
+ \alpha _3^{(1)} \alpha _3^{(2)}  = 0
\end{equation}

Finally, for $E_3=i\lambda_3$ we obtain two equations

\begin{equation}\label{56}
\begin{gathered}
  \alpha _1^{(3)}  + e^{ - ikd} \alpha _2^{(3)}  + \alpha _3^{(3)}  = 0 \hfill \\
  \alpha _1^{(3)}  + e^{ikd} \alpha _2^{(3)}  + \alpha _3^{(3)}  = 0 \hfill \\
\end{gathered}
\end{equation}

which provide with account for the normalizing condition
(\ref{51a}) the following result:

\begin{equation}\label{57}
\alpha _1^{(3)}  =  \pm \frac{1} {{\sqrt 2 }};\alpha _3^{(3)}  =
\mp \frac{1} {{\sqrt 2 }};\alpha _2^{(3)}  = 0;\;
\end{equation}

Therefore, the coefficients of collective states $\left| {\Psi _1
(t)} \right\rangle$  and $\left| {\Psi _2 (t)} \right\rangle$
depend on $kd$, while those of $\left| {\Psi _3 (t)}
\right\rangle$ are $kd$- independent.

Below, we consider special cases. For $kd=2\pi n$, where $n$ is
integer, we obtain from (\ref{53a}), (\ref{53b}) $\alpha _1^{(1)}
= \alpha _2^{(1)}  = \alpha _3^{(1)}  =  \pm \frac{1} {{\sqrt 3
}}$. However, the calculation of the coefficients $\alpha
_n^{(2)}$ from (\ref{54a}), (\ref{54b}) is not straightforward: at
this point both the numerator and the denominator in these
equations are equal to zero. In order to resolve this uncertainty
we put in these equations $kd=2\pi n+\epsilon$, where $\epsilon$
tends to zero. In this case, both the numerator and the
denominator tend to zero as $\epsilon$, and their ratio is finite.
The calculations show that as $kd$ tends to $2\pi n$ the
coefficients  $\alpha _n^{(2)}$ tend to their finite values:
$\alpha _1^{(2)}  = \alpha _3^{(2)}  = \frac{1} {{\sqrt 6
}};\;\alpha _2^{(2)}  =  - \frac{2} {{\sqrt 6 }}$.

For $kd=(2n+1)\pi$ the picture is vise versa: $\alpha _1^{(1)}  =
\alpha _3^{(1)}  = \frac{1} {{\sqrt 6 }};\;\alpha _2^{(1)}  =  -
\frac{2} {{\sqrt 6 }}$, while $\alpha _1^{(2)}  = \alpha _2^{(2)}
= \alpha _3^{(2)}  =  \pm \frac{1} {{\sqrt 3 }}$.

For $kd=(2n+1)\pi /2$ we obtain from (\ref{53a}), (\ref{53b}), and
(\ref{54a}), (\ref{54b}):

\begin{equation}\label{58a}
\begin{gathered}
  \alpha _1^{(1)}  = \alpha _3^{(1)}  = \frac{{ - {\text{i}} + ( - 1)^n \sqrt 7 }}
{{\sqrt 2 \left( {7 - {\text{i}}( - 1)^n 5\sqrt 7 } \right)^{1/2} }} \hfill \\
  \alpha _2^{(1)}  = \frac{{( - 1)^n 3 - {\text{i}}\sqrt 7 }}
{{\sqrt 2 \left( {7 - {\text{i}}( - 1)^n 5\sqrt 7 } \right)^{1/2} }} \hfill \\
\end{gathered}
\end{equation}

\begin{equation}\label{58b}
\begin{gathered}
  \alpha _1^{(2)}  = \alpha _3^{(2)}  = \frac{{ - {\text{i}} - ( - 1)^n \sqrt 7 }}
{{\sqrt 2 \left( {7 + i( - 1)^n 5\sqrt 7 } \right)^{1/2} }} \hfill \\
  \alpha _2^{(2)}  = \frac{{( - 1)^n 3 + i\sqrt 7 }}
{{\sqrt 2 \left( {7 + i( - 1)^n 5\sqrt 7 } \right)^{1/2} }} \hfill \\
\end{gathered}
\end{equation}

Below we summarize these results in the explicit forms of
collective wave functions for above special cases. If $kd$ is
equal to integer multiple of $\pi$ we obtain:

\begin{equation}\label{59a}
\begin{gathered}
  \left| {\Psi _{1`} (t)} \right\rangle  = e^{ - \frac{3}
{2}\Gamma t} \frac{1}
{{\sqrt 3 }}\left( {\left| 1 \right\rangle  + \left| 2 \right\rangle  + \left| 3 \right\rangle } \right) \hfill \\
  \left| {\Psi _{2`} (t)} \right\rangle  = \frac{1}
{{\sqrt 6 }}\left( {\left| 1 \right\rangle  - 2\left| 2 \right\rangle  + \left| 3 \right\rangle } \right) \hfill \\
  \left| {\Psi _{3`} (t)} \right\rangle  = \frac{1}
{{\sqrt 2 }}\left( {\left| 1 \right\rangle  - \left| 3 \right\rangle } \right) \hfill \\
\end{gathered}
\end{equation}

Therefore, two of these collective states $\left| {\Psi _2 (t)}
\right\rangle$  and $\left| {\Psi _3 (t)} \right\rangle$ are dark
as their decay widths are zero.

\vfill

\begin{widetext}
For $kd=(2n+1)\pi /2$ the collective wave functions are as
follows:
\begin{equation}\label{59c}
\begin{gathered}
  \left| {\Psi _{1`} (t)} \right\rangle  = \frac{{e^{ - \frac{1}
{4}\Gamma t} e^{ - i\frac{{\sqrt 7 }} {4}( - 1)^n \Gamma t} }}
{{\sqrt 2 \left( {7 - i( - 1)^n 5\sqrt 7 } \right)^{1/2} }}\left( {( - i + ( - 1)^n \sqrt 7 )\left( {\left| 1 \right\rangle  + \left| 3 \right\rangle } \right) + (( - 1)^n 3 - i\sqrt 7 )\left| 2 \right\rangle } \right) \hfill \\
  \left| {\Psi _{2`} (t)} \right\rangle  =  - \frac{{e^{ - \frac{1}
{4}\Gamma t} e^{i\frac{{\sqrt 7 }} {4}( - 1)^n \Gamma t} }}
{{\sqrt 2 \left( {7 + i( - 1)^n 5\sqrt 7 } \right)^{1/2} }}\left( {(i + ( - 1)^n \sqrt 7 )\left( {\left| 1 \right\rangle  + \left| 3 \right\rangle } \right) - (( - 1)^n 3 + i\sqrt 7 )\left| 2 \right\rangle } \right) \hfill \\
  \left| {\Psi _{3`} (t)} \right\rangle  = e^{ - \Gamma t} \frac{1}
{{\sqrt 2 }}\left( {\left| 1 \right\rangle  - \left| 3 \right\rangle } \right) \hfill \\
\end{gathered}
\end{equation}

where $n$ is any integer.

It is instructive to rewrite (\ref{59c}) in terms of the dark,
$|D\rangle$ and bright, $|B\rangle$ states for a \emph{two qubits}
(the first qubit and the third one) separated by $\lambda/2$. For
$n=0$ we obtain from (\ref{59c}):

\begin{equation}\label{59c1}
\begin{gathered}
  \left| {\Psi _{1`} (t)} \right\rangle  = \frac{{e^{ - \frac{1}
{4}\Gamma t} e^{ - i\frac{{\sqrt 7 }} {4}\Gamma t} }}
{{\sqrt 2 \left( {7 - i5\sqrt 7 } \right)^{1/2} }}\left( {( - i + \sqrt 7 )\sqrt 2 \left| D \right\rangle  \otimes \left| {g_2 } \right\rangle  + (3 - i\sqrt 7 )\left| G \right\rangle  \otimes \left| {e_2 } \right\rangle } \right) \hfill \\
  \left| {\Psi _{2`} (t)} \right\rangle  =  - \frac{{e^{ - \frac{1}
{4}\Gamma t} e^{i\frac{{\sqrt 7 }} {4}\Gamma t} }}
{{\sqrt 2 \left( {7 + i5\sqrt 7 } \right)^{1/2} }}\left( {(i + \sqrt 7 )\sqrt 2 \left| D \right\rangle  \otimes \left| {g_2 } \right\rangle  - (3 + i\sqrt 7 )\left| G \right\rangle  \otimes \left| {e_2 } \right\rangle } \right) \hfill \\
  \left| {\Psi _{3`} (t)} \right\rangle  = e^{ - \Gamma t} \left| B \right\rangle  \otimes \left| {g_2 } \right\rangle  \hfill \\
\end{gathered}
\end{equation}

\end{widetext}

where $|G\rangle=|g_1g_3\rangle$ and

\begin{equation}\label{DB}
\left| D \right\rangle  = \frac{{\left| {e_1 g_3 } \right\rangle +
\left| {g_1 e_3 } \right\rangle }} {{\sqrt 2 }};\;\left| B
\right\rangle  = \frac{{\left| {e_1 g_3 } \right\rangle  - \left|
{g_1 e_3 } \right\rangle }} {{\sqrt 2 }}
\end{equation}

As is seen from the third equation in (\ref{59c1}) the bright
state of the two-qubit system decays independently on the presence
of the second (central) qubit. However, the decay of the dark
state can be revealed only through its entanglement with the
second qubit \cite{Mirho2019}. In fact, the interaction between
second qubit and the dark state formed by two edge qubits gives
rise to the vacuum Rabi oscillations which are shown in
(\ref{eq:math:a34}), (\ref{eq:math:a35}), and in Fig.\ref{Fig4},
Fig.\ref{FIG:Fig5}.

\subsection{Relation between qubits amplitudes and collective
states}

Here, we show how the qubits amplitudes  $\beta_n(t)$ are related
to collective wave functions $\left| {\Psi _i (t)} \right\rangle$
(\ref{45}). We write the dynamic wave function $\left| {\Psi (t)}
\right\rangle$ (\ref{43}) as a decomposition over the collective
states (\ref{45}).

\begin{equation}\label{60}
\left| {\Psi (t)} \right\rangle  = \sum\limits_{i = 1}^3 {A_i
\left| {\Psi _i (t)} \right\rangle }  = \sum\limits_{i,n = 1}^3
{A_i e^{ - i\bar E_i t} \alpha _n^{(i)} \left| n \right\rangle }
\end{equation}

From (\ref{43}) we obtain:
\begin{equation}\label{61}
\beta _n (t) = \sum\limits_{i = 1}^3 {e^{ - i\bar E_i t} A_i
\alpha _n^{(i)} }
\end{equation}

with the initial conditions
\begin{equation}\label{62}
\begin{gathered}
  \beta _{n_0 } (0) = \sum\limits_{i = 1}^3 {A_i \alpha _{n_0 }^{(i)} }  = 1
  \\
  \beta _n (0) = \sum\limits_{i = 1}^3 {A_i \alpha _n^{(i)} }  = 0;\;n \ne n_0
\end{gathered}
\end{equation}

where $n_0$ is the sequence number of excited qubit.

The probability amplitude $|\beta_n(t)|^2$ can directly be
expressed in terms of collective state wave functions
$|\Psi_i(t)\rangle$.

\begin{equation}\label{b2}
\left| {\beta _n (t)} \right|^2  = \left\langle n \right|\left(
{\sum\limits_{i,j = 1}^3 {} A_i A_j^* \left| {\Psi _i }
\right\rangle \left\langle {\Psi _j } \right|} \right)\left| n
\right\rangle
\end{equation}

From linear algebraic equations (\ref{62}) we can find
coefficients $A_i$ and, therefore, restore the qubits' amplitudes
$\beta_n(t)$. However, if the number of qubits is large, this
procedure is not convenient for computer simulations.  The main
reason is that it requires first, the calculation of $N$ complex
energies from determinant of $N\times N$ matrix analogous to
(\ref{48}), second, the calculations of $\alpha_n^{(i)}$ from non
linear conditions (\ref{51a}) and (\ref{51b}), and third, the
solution of a system of $N$ linear algebraic equations analogous
to (\ref{62}). Every of these three steps is not simple from a
mathematical point of view. It is more convenient to directly
compute the qubits amplitudes  $\beta_n(t)$ from a set of the
linear differential equations (\ref{eq:math:a15}), which allow us
to completely avoid all three steps we mentioned above.

Nevertheless, we should like to mention some interesting
consequences that follow from equations (\ref{61}), (\ref{62}),
(\ref{b2}). First, from (\ref{61}) we see that the dark states
($Im\bar E_i  = 0$) contribute to $\beta_n(t)$, even though they
do not contribute to the spectrum of the photon emission. Second,
it follows from (\ref{b2}) that if specific collective state
$|\Psi_i(t)\rangle$ does not contain qubit state $|n\rangle$, then
this collective state does not take part in the formation of the
dynamics of the qubits amplitudes $\beta_n(t)$. As was shown above
(see (\ref{57})) the state $|\Psi_3(t)\rangle$ does not contain
the qubit state $|2\rangle$. Therefore, independently on the value
of $kd$, only two collective states $\left| {\Psi _1 (t)}
\right\rangle$ and $\left| {\Psi _2 (t)} \right\rangle$ take part
in the formation of the dynamics of the qubits amplitudes
$\beta_2(t)$.

\section{Three non-identical qubits}

Here, we consider a system in which all three qubits are identical
except for the frequency of a second qubit which has a different
value $\Omega_0$. For this case, we rewrite the equations
(\ref{eq:math:a11}), (\ref{eq:math:a12}), and (\ref{eq:math:a13}).

\begin{equation}\label{64a}
\begin{gathered}
  \frac{{d\beta _1 }}
{{dt}} =  - \frac{\Gamma } {2}\beta _1 (t) - i\frac{{\delta \Omega
}}
{2}\beta _1 (t) \hfill \\
   - \beta _2 (t)\frac{1}
{2}\left( {\frac{{\Omega _0 }} {\Omega }} \right)^{1/2} \Gamma
e^{ik_0 d}  - \frac{\Gamma }
{2}\beta _3 (t)e^{2ikd}  \hfill \\
\end{gathered}
\end{equation}

\begin{equation}\label{64b}
\begin{gathered}
  \frac{{d\beta _2 }}
{{dt}} =  - \frac{\Gamma } {2}\beta _2 (t) + i\frac{{\delta \Omega
}}
{2}\beta _2 (t) \hfill \\
   - \frac{1}
{2}\left( {\frac{\Omega }
{{\Omega _0 }}} \right)^{1/2} \Gamma e^{ikd} \left( {\beta _1 (t) + \beta _3 (t)} \right) \hfill \\
\end{gathered}
\end{equation}

\begin{equation}\label{64c}
\begin{gathered}
  \frac{{d\beta _3 }}
{{dt}} =  - \frac{\Gamma } {2}\beta _3 (t) - i\frac{{\delta \Omega
}} {2}\beta _3 (t) - \beta _1 (t)\frac{\Gamma }
{2}e^{2ikd}  \hfill \\
   - \beta _2 (t)\frac{1}
{2}\left( {\frac{{\Omega _0 }}
{\Omega }} \right)^{1/2} \Gamma e^{ik_0 d}  \hfill \\
\end{gathered}
\end{equation}

where $\delta \Omega  = \Omega  - \Omega _0$, $k = \frac{\Omega }
{{v_g }},\;k_0  = \frac{{\Omega _0 }} {{v_g }}$

Characteristic roots can be found from the determinant of
equations (\ref{64a}), (\ref{64b}), and (\ref{64c}).

\begin{equation}\label{65a}
\begin{gathered}
  \lambda _1  =  - \frac{\Gamma }
{2}\left( {1 + \frac{1}
{2}e^{2ikd} } \right) \hfill \\
   - \frac{\Gamma }
{4}e^{ikd} \sqrt {e^{2ikd}  + 8e^{i(k_0  - k)d}  + 4i\frac{{\delta
\Omega }} {\Gamma } - 4\left( {\frac{{\delta \Omega }}
{\Gamma }} \right)^2 e^{ - 2ikd} }  \hfill \\
\end{gathered}
\end{equation}

\begin{equation}\label{65b}
\begin{gathered}
  \lambda _2  =  - \frac{\Gamma }
{2}\left( {1 + \frac{1}
{2}e^{2ikd} } \right) \hfill \\
   + \frac{\Gamma }
{4}e^{ikd} \sqrt {e^{2ikd}  + 8e^{i(k_0  - k)d}  + 4i\frac{{\delta
\Omega }} {\Gamma } - 4\left( {\frac{{\delta \Omega }}
{\Gamma }} \right)^2 e^{ - 2ikd} }  \hfill \\
\end{gathered}
\end{equation}

\begin{equation}\label{65c}
\lambda _3  = \frac{\Gamma } {2}e^{i2kd}  - \frac{\Gamma } {2} -
i\frac{{\delta \Omega }} {2}
\end{equation}

The calculations show that equations (\ref{65a}) and (\ref{65b})
provide no dark states $(\operatorname{Re} (\lambda _{1,2} ) = 0)$
if  $\delta\Omega$ is not equal to zero. The quantities
$\operatorname{Re} (\lambda _{1,2} )$  are always negative at any
$kd$. The exception is the third root $\lambda_3$, with real part
being equal to zero for $kd=n\pi$. Therefore, if the root
$\lambda_3$ does not contribute to the dynamics of qubits, then at
any $kd$, the qubits' amplitudes will decay to zero with the rate
being dependent on $\delta\Omega$.

\begin{figure} [h]
  \includegraphics[width=8 cm]{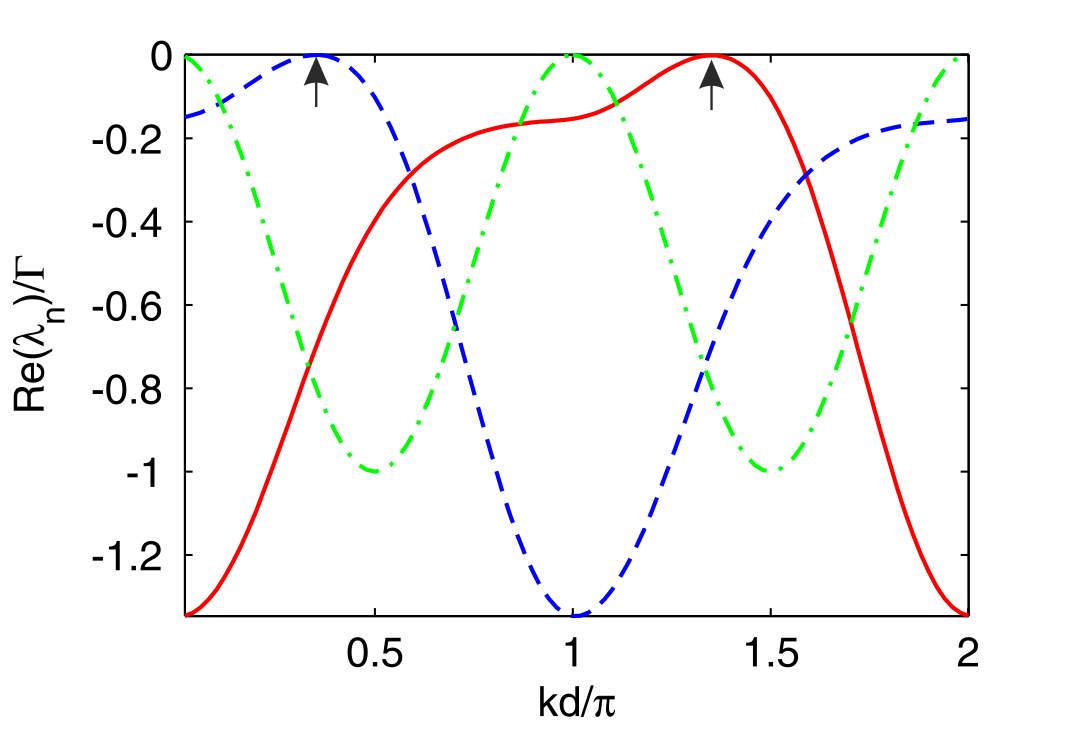}\\
  \caption{Dependence of $\operatorname{Re} (\lambda _n )/\Gamma$
 on $kd$ for $\delta\Omega=\Gamma$ . Solid (red) line relates to $\lambda_1$, dashed (blue)
  line relates to  $\lambda_2$, and dashed-dotted (green) line relates to  $\lambda_3$.
  The arrows show the points of deep subradiant states where
  $\operatorname{Re} (\lambda _{1,2} )/\Gamma  \approx  - 0.001$ }\label{Fig7}
\end{figure}

The dependence of real parts of $\lambda_n$  on $kd$ is shown in
Fig.\ref{Fig7} for $\delta\Omega=\Gamma$.

If the second (central) qubit is excited, the amplitudes of the
first and the third qubit are the same $\beta _1 (t) = \beta _3
(t) \equiv \beta (t)$ . Then three equations (\ref{64a}),
(\ref{64b}), and (\ref{64c}) reduce to two equations, for
$\beta_2$ and $\beta$.

\begin{equation}\label{66a}
\frac{{d\beta }} {{dt}} =  - \frac{\Gamma } {2}\left( {1 +
e^{2ikd}  + i\frac{{\delta \Omega }} {\Gamma }} \right)\beta (t) -
\beta _2 (t)\frac{\Gamma } {2}\left( {\frac{{\Omega _0 }} {\Omega
}} \right)^{1/2} e^{ik_0 d}
\end{equation}

\begin{equation}\label{66b}
\frac{{d\beta _2 }} {{dt}} =  - \frac{\Gamma } {2}\left( {1 -
i\frac{{\delta \Omega }} {\Gamma }} \right)\beta _2 (t) - \beta
(t)\Gamma \left( {\frac{\Omega } {{\Omega _0 }}} \right)^{1/2}
e^{ikd}
\end{equation}

The characteristic roots of these equations are equal to
$\lambda_1$ and $\lambda_2$  given above in equations (\ref{65a})
and (\ref{65b}).

The solution of equations (\ref{66a}) and (\ref{66b}). with
account for initial conditions for the amplitudes, $\beta _2 (0) =
1,\;\beta (0) = 0$  and their time derivatives

\begin{equation}\label{67a}
\left. {\frac{{d\beta }} {{dt}}} \right|_{t = 0}  =  -
\frac{\Gamma } {2}\left( {\frac{{\Omega _0 }} {\Omega }}
\right)^{1/2} e^{ik_0 d}
\end{equation}

\begin{equation}\label{67b}
\left. {\frac{{d\beta _2 }} {{dt}}} \right|_{t = 0}  =  -
\frac{\Gamma } {2}\left( {1 - i\frac{{\delta \Omega }} {\Gamma }}
\right)
\end{equation}
is similar to (\ref{eq:math:a23})
\begin{equation}\label{23}
\begin{gathered}
  \beta _2 (t) = b_1 e^{\lambda _1 t}  + b_2 e^{\lambda _2 t}  \hfill \\
  \beta (t) = a_1 e^{\lambda _1 t}  + a_2 e^{\lambda _2 t}  \hfill \\
\end{gathered}
\end{equation}
with

\begin{equation}\label{68}
\begin{gathered}
  b_1  =  - \frac{{\frac{\Gamma }
{2}\left( {1 - i\frac{{\delta \Omega }} {\Gamma }} \right) +
\lambda _2 }}
{{\lambda _1  - \lambda _2 }} \hfill \\
  b_2  = \frac{{\frac{\Gamma }
{2}\left( {1 - i\frac{{\delta \Omega }} {\Gamma }} \right) +
\lambda _1 }}
{{\lambda _1  - \lambda _2 }} \hfill \\
\end{gathered}
\end{equation}

\begin{equation}\label{69}
\begin{gathered}
  a_1  =  - \frac{\Gamma }
{2}\left( {\frac{{\Omega _0 }} {\Omega }} \right)^{1/2}
\frac{{e^{ik_0 d} }}
{{\lambda _1  - \lambda _2 }} \hfill \\
  a_2  = \frac{\Gamma }
{2}\left( {\frac{{\Omega _0 }} {\Omega }} \right)^{1/2}
\frac{{e^{ik_0 d} }}
{{\lambda _1  - \lambda _2 }} \hfill \\
\end{gathered}
\end{equation}

\begin{figure}[h]
  \includegraphics[width=8 cm]{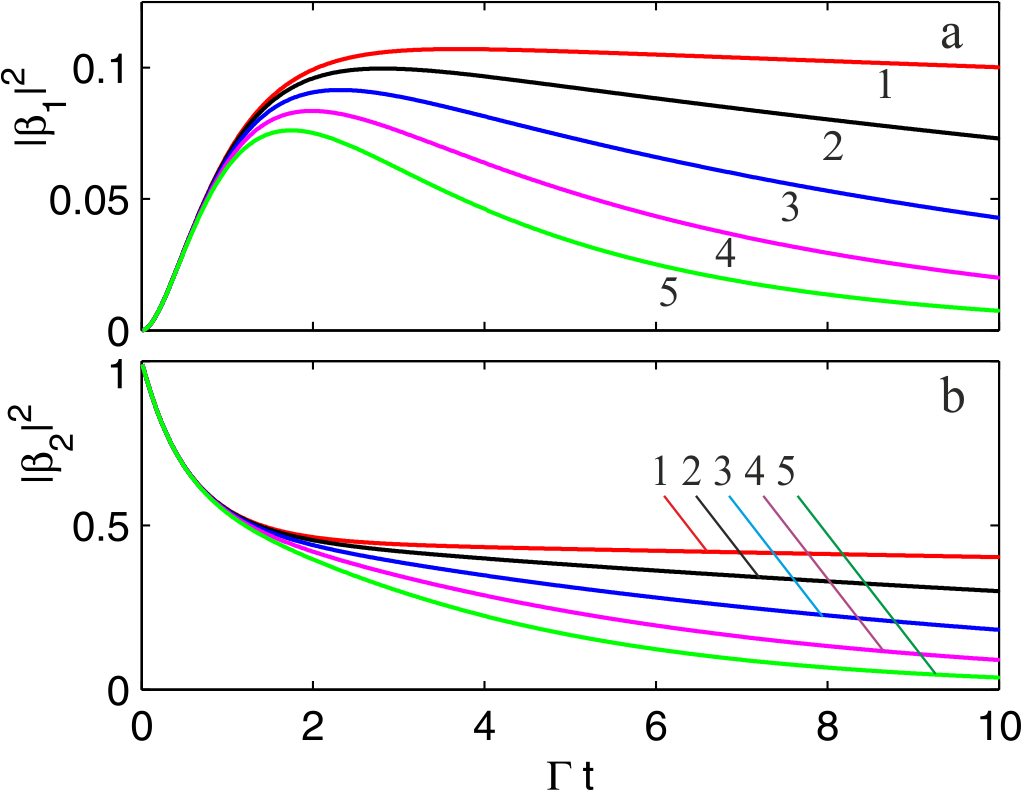}\\
   \caption{Dependence of the decay rates on the detuning  $\delta\Omega$
   for initially excited second qubit. $kd=2\pi$ . (a)-the decay rates of the
   first(third)
   qubit; (b) the decay rates of the second qubit. The numbers in the plots
   correspond to different detunings:
    $1- \delta\Omega=0.2\Gamma ;\, 2- \delta\Omega=0.4\Gamma ; \,3- \delta\Omega=0.6\Gamma ;
    \, 4- \delta\Omega=0.8\Gamma ;\, 5- \delta\Omega=\Gamma$.}\label{Fig8}
\end{figure}

Using the detuning  $\delta\Omega$ as  the external parameter we
can control the decay rates of the qubits amplitudes. This is
shown in Fig.\ref{Fig8} for several values of the detuning
$\delta\Omega$ with the second qubit being initially excited.  In
the case shown in Fig.\ref{Fig8}, the decay rates are governed by
real parts of two roots, $\lambda_1$ and $\lambda_2$ with
$|Re\lambda_1| \ll |Re\lambda_2|$. As $\delta\Omega/\Gamma$
increases from $0$ to $1$, $|Re\lambda_1|$  decreases from $1.5$
to $1.346$ while $|Re\lambda_2|$ increases from $0$ to $0.154$.
Even though the first root tends to slow down the decay rate as
$\delta\Omega/\Gamma$ increases, its influence for $t\gg\Gamma$
becomes negligible, so that the main contribution to the decay
rate for large times comes from the second root which speeds up
the decay rate as $\delta\Omega/\Gamma$ increases.

Therefore, the greater is $\delta\Omega$, the more is its
influence on the qubit decay rates.

If the first qubit is initially excited the solution is similar to
(\ref{eq:math:a30}):

\begin{equation}\label{70}
\begin{gathered}
  \beta _1 (t) = \frac{b_2}{2}e^{\lambda_1 t}
  + \frac{b_1}{2}e^{\lambda_2 t}
+ \frac{1}{2}e^{_{\lambda _3 t} }  \hfill \\
  \beta _2 (t) = a_1 \left( e^{\lambda_1 t}  - e^{\lambda_2 t} \right) \hfill \\
 \beta _3 (t) = \frac{b_2}{2}e^{\lambda_1 t}
  + \frac{b_1}{2}e^{\lambda_2 t}
-\frac{1}{2}e^{_{\lambda _3 t} }
\end{gathered}
\end{equation}

where $\lambda_1$,  $\lambda_2$,  $\lambda_3$, and  $a_1$, $b_1$,
$b_2$ are given in (\ref{65a}), (\ref{65b}), (\ref{65c}), and
(\ref{68}), (\ref{69}).

For this case, we show in Fig. 9 the temporal behavior of qubits
amplitudes for several values of the detuning  $\delta\Omega$. In
the absence of detuning (the panel \emph{a}) we obtain the result
shown in Fig.\ref{FIG:Fig3}. However, for non zero detuning the
amplitudes of the first and the third qubits decay to 0.25, while
the second qubit decays to zero. This difference can be explained
by the influence of dark states (the last terms in the expressions
for $\beta_(t)$ and $\beta_3(t)$ in (\ref{70})). In general, as
the detuning increases the decay rates increase as well.

\begin{figure} [h]
  \includegraphics[width=8 cm]{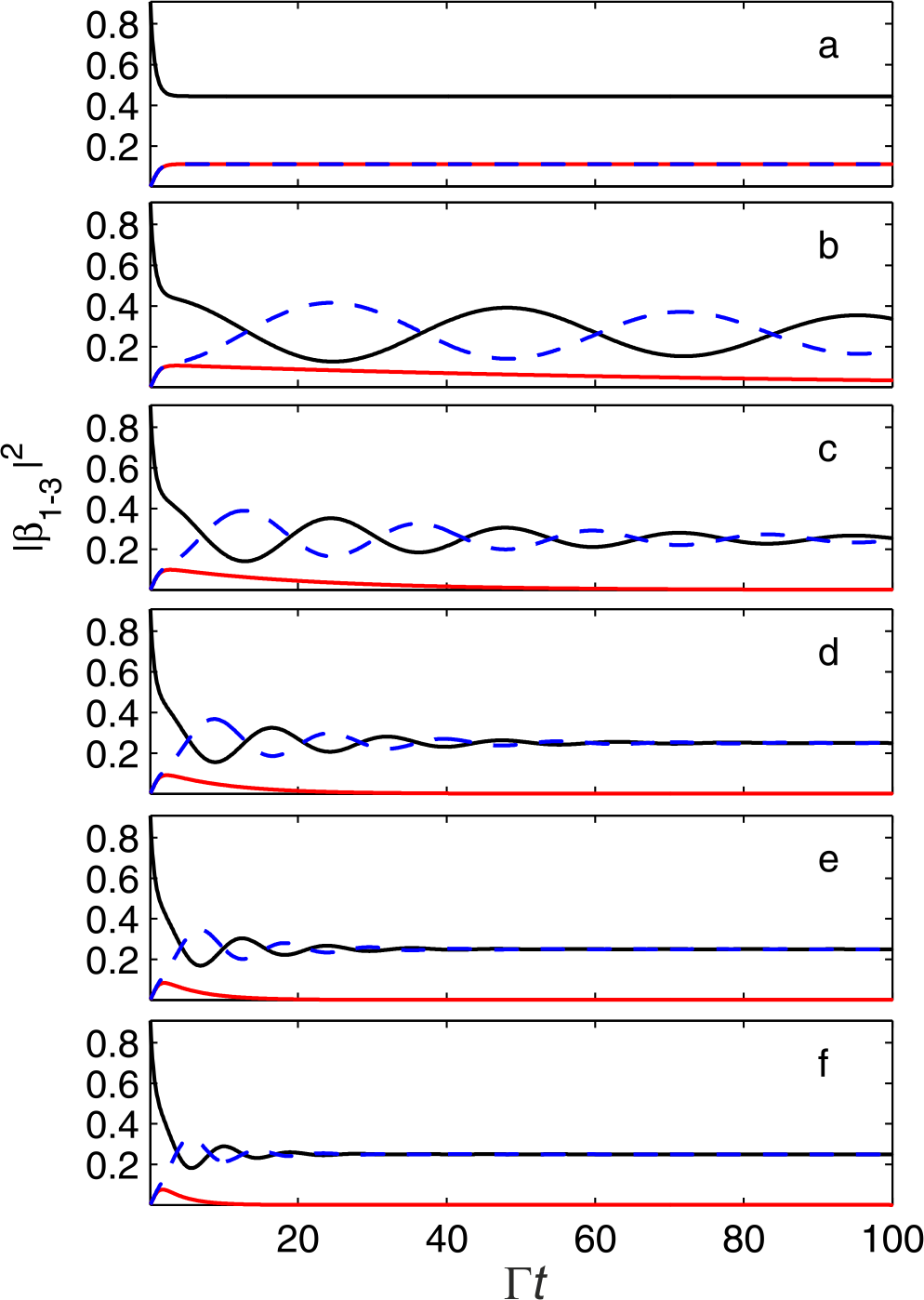}\\
  \caption{Dependence of the decay rates on the detuning  $\delta\Omega$
   for initially excited the first qubit. $kd=2\pi$, $\Gamma/\Omega=0.001$.
   Every panel from \emph{a} to \emph{f} shows the decay rates of the
   the first (black, solid line), second (red, solid line), and third
   (blue, dashed line) qubit, respectively.  From top (\emph{a}) to bottom (\emph{f}) the quantity
  $\delta\Omega/\Gamma$ is varied from $0$ (\emph{a})
 to $1$ (\emph{f}) with the increment being equal to $0.2$}\label{Fig9}
\end{figure}

The subsystem consisting of a central qubit and the symmetric
qubit array can be described as an analogue to a cavity QED
system. In this description, the central qubit acts as a two-level
atom and the symmetric qubit array mimics a high-finesse cavity,
with the qubits array acting as an atomic mirror \cite{Chang2012}.
It seems that this analogy is supported by Fig.\ref{FIG:Fig3}
where two dark states (\ref{59a}) of the \emph{whole} system
prevent the qubit amplitudes from the damping. If this analogy
worked we would expect that the detuning of a central qubit would
not lead to any damping at all. In real cavity the qubit which is
large detuned from the cavity resonance cannot exchange its energy
with the cavity via real photons. However, as is seen from
Fig.\ref{Fig8}, a full analogy between resonance cavity and the
qubit array does not exist. Even a small detuning of the central
qubit from the frequency of the qubit array  results in the
disruption of the dark states that, in turn, leads to the damping
of the qubits amplitudes.

\subsection{Spectral density of spontaneous photon emission}

For the calculation of the spectral density of photon emission we
use the equation (\ref{eq:math:a9}) together with the substitution
of $\beta_n(t)$ from equations (\ref{eq:math:a14}).

\begin{equation}\label{71}
\begin{gathered}
  \gamma _k (t) =  - ig_k e^{ikd} \int\limits_0^t {} \bar \beta _1 (t')e^{i(\Omega  - \Omega _0 )t'/2} e^{i(\omega  - \Omega )t'} dt' \hfill \\
   - ig_k \int\limits_0^t {} \bar \beta _2 (t')e^{ - i(\Omega  - \Omega _0 )t'/2} e^{i(\omega  - \Omega _0 )t'} dt' \hfill \\
   - ig_k e^{ - ikd} \int\limits_0^t {} \bar \beta _3 (t')e^{i(\Omega  - \Omega _0 )t'/2} e^{i(\omega  - \Omega )t'} dt' \hfill \\
\end{gathered}
\end{equation}
where $ kd = \frac{\omega } {{v_g }}d$, $ g_k^{}  = \left(
{\frac{{v_g \Gamma }} {{2L}}} \right)^{1/2}$.

The amplitudes $\bar\beta_n(t)$ in (\ref{71}) are just the
amplitudes $\beta_n(t)$ given above in (\ref{23}) and (\ref{70}).

\begin{figure}[h]
  \includegraphics[width=8 cm]{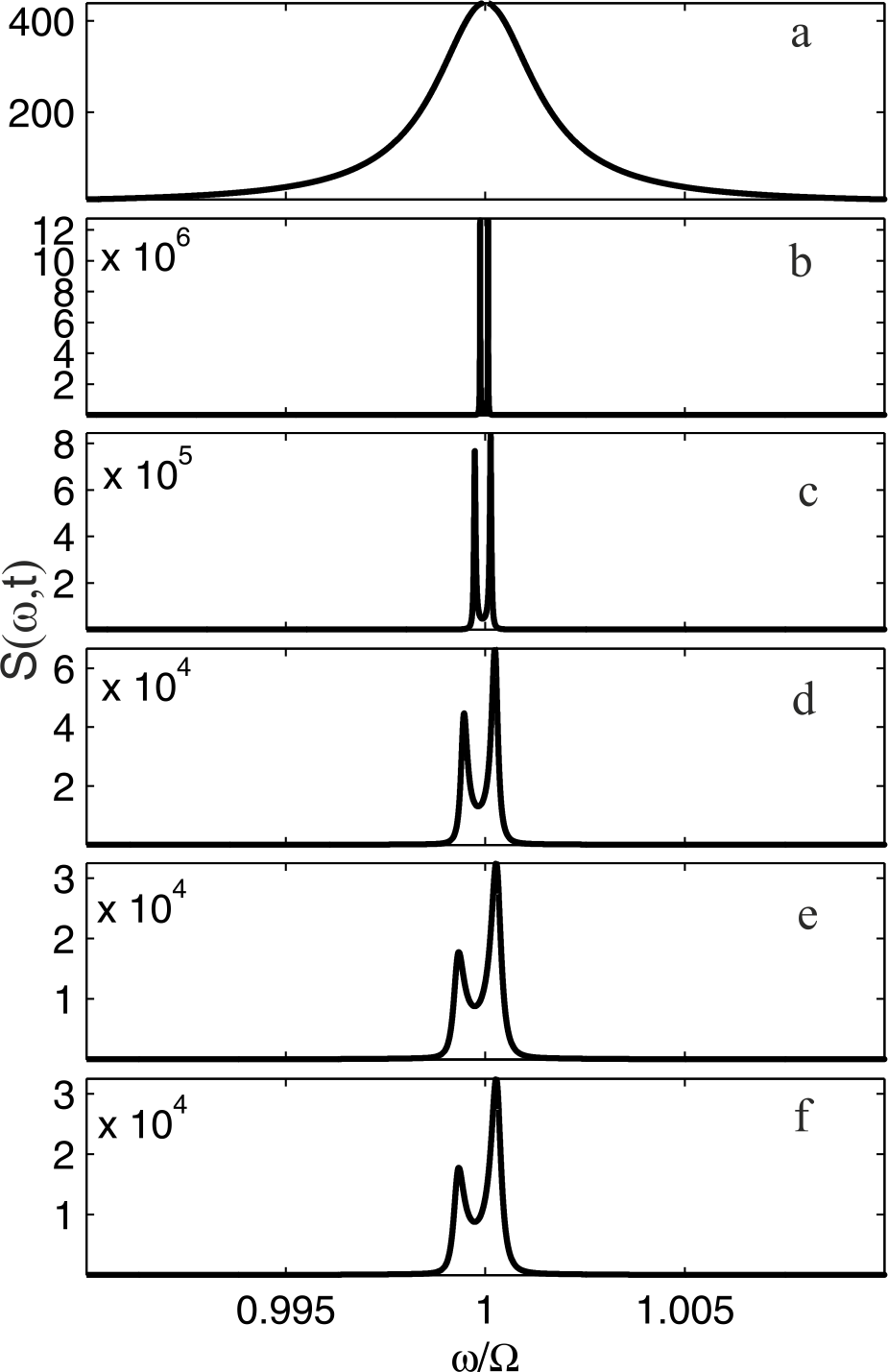}\\
  \caption{Dependence of spectral density on the detuning. Second qubit is initially
  excited. $kd=2\pi,\Gamma/\Omega=0.001$, $t\gg 1/\Gamma$. From top (a) to bottom (f) the quantity
  $\delta\Omega/\Gamma$ is varied from $0$ (a) to $1$ (f)
 with the increment being equal to $0.2$.  }\label{Fig10}
\end{figure}

\begin{figure}[h]
  \includegraphics[width=8 cm]{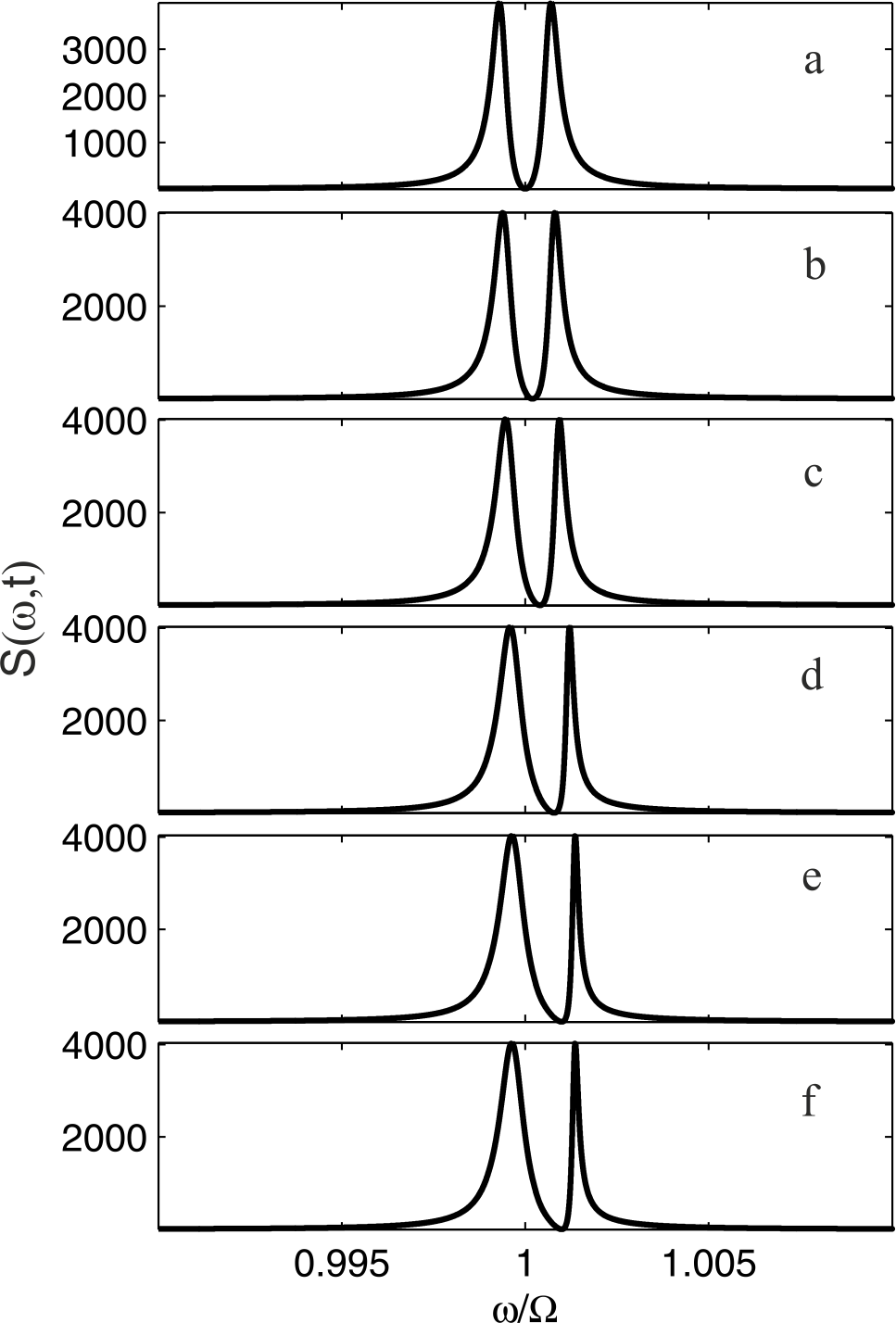}\\
  \caption{Dependence of spectral density on the detuning. Second qubit is initially
  excited. $kd=3\pi/2,\Gamma/\Omega=0.001$, $t\gg 1/\Gamma$. From top (a) to bottom (f) the quantity
  $\delta\Omega/\Gamma$ is varied from $0$ (a) to $1$ (f)
with the increment being equal to $0.2$}\label{Fig11}
\end{figure}

Below we show two plots which demonstrate how the spectral density
of photon emission depends on the detuning $\delta\Omega$. The
dependencies of spectral density on detuning are shown in
Fig.\ref{Fig10} for $kd=2\pi$ and in Fig.\ref{Fig11} for
$kd=3\pi/2$, respectively. In both cases the second qubit was
initially excited.

If $\delta\Omega=0$ and $kd=2\pi$ we see a single Lorentzian peak
(Fig.\ref{Fig10}(a)) with the width equal to $3\Gamma$ (see also
(\ref{eq:math:a41}). If $\delta\Omega\neq 0$ a Lorentzian peak
splits into two narrow peaks. The distance between the peaks,
$\Delta\omega$ is approximately equal to
$(|Im\lambda_1|+|Im\lambda_2|)$. The width of the peaks is
determined by the root with a lowest real part. In the case shown
in Fig.\ref{Fig10}(b-f) the width of the peaks and their height
are approximately equal to $|Re\lambda_2|$ and
$\Omega/|Re\lambda_2|$, respectively. To estimate the order of
these quantities we have found numerically the range of variation
for $Re\lambda_2$, $Im\lambda_1$, and $Im\lambda_2$. As
$\delta\Omega/\Gamma$ increases from $0.2$ to $1$, the quantities
$Re\lambda_2$, $Im\lambda_1$, and $Im\lambda_2$ vary from
$-0.006\Gamma$ to $-0.154\Gamma$, from $+0.028\Gamma$ to
$+0.202\Gamma$, and from $-0.032\Gamma$ to $-0.205\Gamma$,
respectively.

If $\delta\Omega=0$ and $kd=3\pi/2$ there are two similar peaks
shown in Fig.\ref{Fig11}(a) (see also Fig.\ref{FIG6}(a)). The
distance between the peaks and their widths are determined by the
roots $\lambda_1$ and $\lambda_2$:
$Im\lambda_1$=$-Im\lambda_2$=$(\sqrt{7}/4)\Gamma$,
$Re\lambda_1$=$Re\lambda_2$=$-0.25\Gamma$ (see Eq.
(\ref{eq:math:a18})). As $\delta\Omega/\Gamma$ increases from
$0.2$ to $1$, the quantities $Re\lambda_1$ and $Re\lambda_2$ are
of the same order of magnitude: $Re\lambda_1$ varies from
$-0.287\Gamma$ to $-0.399\Gamma$, and $Re\lambda_2$ varies from
$-0.213\Gamma$ to $-0.100\Gamma$. This behavior is confirmed in
Fig.\ref{Fig11} (b-f) where the width of the left peak increases,
while the width of the right peak decreases. The distance between
the peaks are determined by the quantities $Im\lambda_1$ and
$Im\lambda_2$ which vary from $-0.708\Gamma$ to $-0.890\Gamma$,
and from $+0.638\Gamma$ to $+0.821\Gamma$, respectively.

\begin{figure}
  \includegraphics[width=8 cm]{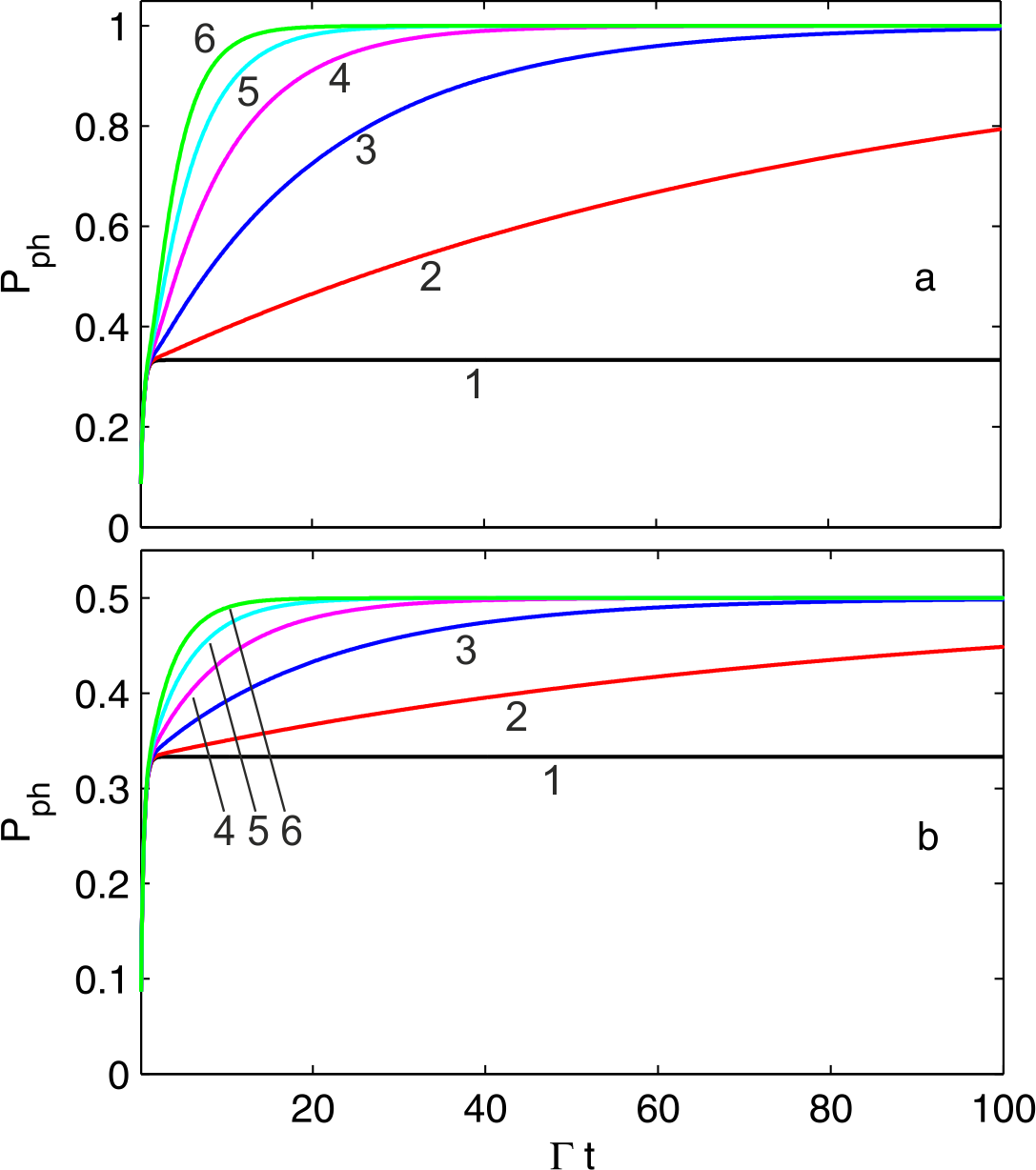}\\
  \caption{The probability amplitude of the photon emission.
 $kd=2\pi,\Gamma/\Omega=0.001$. (a) the second qubit is initially excited;
(b) the first qubit is initially excited. The numbers in the plots
   correspond to different detunings: $1-\delta\Omega=0$,
    $2- \delta\Omega=0.2\Gamma ;\, 3- \delta\Omega=0.4\Gamma ; \,4- \delta\Omega=0.6\Gamma ;
    \, 5- \delta\Omega=0.8\Gamma ;\, 6- \delta\Omega=\Gamma$.}\label{Fig12P}
\end{figure}

We also calculated the temporal behavior of the full probability
of the photon emission , $P_{ph}(t)$ as  a function of the
detuning. This dependence is shown in Fig.\ref{Fig12P}. It is
evident from the definition (\ref{eq:math:a7}) that $P_{ph}(0)=0$
and as the time increases it approaches either $1$ if all qubits
amplitudes damp out to zero, or a constant value if not all qubits
decay to zero. Because the detuning influences the rate of the
qubits damping, the output rate of the photon emission,
$dP_{ph}/dt$ also depends on the detuning. The greater is the
detuning, the greater is the rate of the photon emission.

\section{Conclusion}

In this paper we have thoroughly investigated the dynamics
behavior of qubits amplitudes in 1D chain consisting of three
qubits embedded in an open waveguide. Within a single-excitation
subspace we have found the evolution of qubits amplitudes if one
of the qubits was initially excited. We have shown that even
though the dark states do not contribute to the output photon
emission, they influence the evolution of qubits amplitudes in
that they prevent the qubits amplitudes from decaying to zero. We
have found the collective eigenstates of a three-qubit system and
have shown how the qubits amplitudes can be expressed in terms of
the amplitudes of the collective states. We also calculated the
spectral density of the output photon emission from three-qubit
system and studied its dependence on the $kd$ value. We studied
the case when the frequency of the second qubit was different from
that of the edge qubits. In this case, the rates of qubits decay
crucially depend on the frequency detuning between central and
edge qubits. The greater is the detuning, the greater is the rate
of the qubits damping.

We hope that this research will prove useful for the development
of the efficient control and readout protocols for a few-qubit
quantum processor.

\begin{acknowledgments}
Ya. S. G. thanks A. Sultanov for fruitful discussions. The work is
supported by the Ministry of Education and Science of Russian
Federation under the project FSUN-2020-0004.
\end{acknowledgments}

\appendix

\section{Equations for qubits amplitudes in the Wigner-Weisskopf
approximation}

We assume that the first and the third qubits are identical
$(\Omega _1  = \Omega _3  \equiv \Omega ,\;g_k^{(1)}  = g_k^{(3)}
\equiv g_k)$ The frequency and the coupling of the second qubit
are different ($ \Omega _2  \equiv \Omega _0 ,\,g_k^{(2)} \equiv
g_k^{(0)}$). A distance between central qubit and the edge qubits
is equal to $d$. We take the origin in the location of the second
qubit: $x_1= -d, x_2=0, x_3=+d$. For this case, we expand the
equation (\ref{eq:math:a8}) as a set of three equations.

\begin{equation}\label{A1}
\begin{gathered}
  \frac{{d\beta _1 }}
{{dt}} =  - \sum\limits_k {g_k^2 } \int\limits_0^t {\beta _1 (t')e^{ - i(\omega _k  - \Omega )(t - t')} dt'}  \hfill \\
   - \sum\limits_k {g_k^{} g_k^{(0)} e^{ - ikd} e^{i(\Omega  - \Omega _0 )t} } \int\limits_0^t {\beta _2 (t')e^{ - i(\omega _k  - \Omega _0 )(t - t')} dt'}  \hfill \\
   - \sum\limits_k {g_k^2 e^{ - i2kd} } \int\limits_0^t {\beta _3 (t')e^{ - i(\omega _k  - \Omega )(t - t')} dt'}  \hfill \\
\end{gathered}
\end{equation}

\begin{equation}\label{A2}
\begin{gathered}
  \frac{{d\beta _2 }}
{{dt}} =  - \sum\limits_k {\left( {g_k^{(0)} } \right)^2 } \int\limits_0^t {\beta _2 (t')e^{ - i(\omega _k  - \Omega _0 )(t - t')} dt'}  \hfill \\
   - \sum\limits_k {g_k^{(0)} g_k^{} e^{ikd} e^{i(\Omega _0  - \Omega )t} } \int\limits_0^t {\beta _1 (t')e^{ - i(\omega _k  - \Omega )(t - t')} dt'}  \hfill \\
   - \sum\limits_k {g_k^{(0)} g_k^{} e^{ - ikd} e^{i(\Omega _0  - \Omega )t} } \int\limits_0^t {\beta _3 (t')e^{ - i(\omega _k  - \Omega )(t - t')} dt'}  \hfill \\
\end{gathered}
\end{equation}

\begin{equation}\label{A3}
\begin{gathered}
  \frac{{d\beta _3 }}
{{dt}} =  - \sum\limits_k {g_k^2 } \int\limits_0^t {\beta _3 (t')e^{ - i(\omega _k  - \Omega )(t - t')} dt'}  \hfill \\
   - \sum\limits_k {g_k^2 e^{i2kd} } \int\limits_0^t {\beta _1 (t')e^{ - i(\omega _k  - \Omega _{} )(t - t')} dt'}  \hfill \\
   - \sum\limits_k {g_k^{} g_k^{(0)} e^{ikd} e^{i(\Omega  - \Omega _0 )t} } \int\limits_0^t {\beta _2 (t')e^{ - i(\omega _k  - \Omega _0 )(t - t')} dt'}  \hfill \\
\end{gathered}
\end{equation}

According to Wigner-Weisskopf approach we replace $\beta _1
(t'),\beta _2 (t'),\beta _3 (t')$ in the integrands with $\beta _1
(t),\beta _2 (t),\beta _3 (t)$ and take them out of the integrals.

\begin{equation}\label{A4}
\begin{gathered}
  \frac{{d\beta _1 }}
{{dt}} =  - \beta _1 (t)\sum\limits_k {g_k^2 } I_k (\Omega ,t) \hfill \\
   - \beta _2 (t)\sum\limits_k {g_k^{} g_k^{(0)} e^{ - ikd} e^{i(\Omega  - \Omega _0 )t} } I_k (\Omega _0 ,t) \hfill \\
   - \beta _3 (t)\sum\limits_k {g_k^2 e^{ - i2kd} } I_k (\Omega ,t) \hfill \\
\end{gathered}
\end{equation}

\begin{equation}\label{A5}
\begin{gathered}
  \frac{{d\beta _2 }}
{{dt}} =  - \beta _2 (t)\sum\limits_k {|g_k^{(0)} |^2 } I_k (\Omega _0 ,t) \hfill \\
   - \beta _1 (t)\sum\limits_k {g_k^{(0)} g_k^{} e^{ikd} e^{ - i(\Omega  - \Omega _0 )t} } I_k (\Omega ,t) \hfill \\
   - \beta _3 (t)\sum\limits_k {g_k^{(0)} g_k^{} e^{ - ikd} e^{ - i(\Omega  - \Omega _0 )t} } I_k (\Omega ,t) \hfill \\
\end{gathered}
\end{equation}

\begin{equation}\label{A6}
\begin{gathered}
  \frac{{d\beta _3 }}
{{dt}} =  - \beta _3 (t)\sum\limits_k {g_k^2 } I_k (\Omega ,t) \hfill \\
   - \beta _1 (t)\sum\limits_k {g_k^2 e^{i2kd} } I_k (\Omega ,t) \hfill \\
   - \beta _2 (t)\sum\limits_k {g_k^{} g_k^{(0)} e^{ikd} e^{i(\Omega  - \Omega _0 )t} } I_k (\Omega _0 ,t) \hfill \\
\end{gathered}
\end{equation}
where
\begin{equation}\label{A7}
\begin{gathered}
  I_k (\Omega ,t) = \int\limits_0^t {e^{ - i(\omega _k  - \Omega )(t - t')} dt'}  = \int\limits_0^t {e^{ - i(\omega _k  - \Omega )\tau } d\tau }  \hfill \\
   \approx \int\limits_0^\infty  {e^{ - i(\omega _k  - \Omega )\tau } d\tau }  = \pi \delta (\omega _k  - \Omega ) - iP.v.\left( {\frac{1}
{{\omega _k  - \Omega }}} \right) \hfill \\
\end{gathered}
\end{equation}
where $P.v.$ is a Cauchy principal value integral.

We can remove oscillating exponents  in (\ref{A4})-(\ref{A6}) with
the aid of the substitution

\begin{equation}\label{A8}
\begin{gathered}
  \beta _1 (t) = e^{i(\Omega  - \Omega _0 )t/2} \bar \beta _1 (t) \hfill \\
  \beta _2 (t) = e^{ - i(\Omega  - \Omega _0 )t/2} \bar \beta _2 (t) \hfill \\
  \beta _3 (t) = e^{i(\Omega  - \Omega _0 )t/2} \bar \beta _3 (t) \hfill \\
\end{gathered}
\end{equation}

In addition, we assume the coupling constants $g_k$ are even
functions of $k$ ($g_k=g_{-k}$). Then from (\ref{A4})-(\ref{A6})
we obtain

\begin{equation}\label{A9}
\begin{gathered}
  \frac{{d\bar \beta _1 }}
{{dt}} =  - \bar \beta _1 (t)\sum\limits_k {g_k^2 } I_k (\Omega
,t) - i\frac{{\Omega  - \Omega _0 }}
{2}\bar \beta _1 (t) \hfill \\
   - \bar \beta _2 (t)2\sum\limits_{k > 0} {g_k^{} g_k^{(0)} \cos (kd)} I_k (\Omega _0 ,t) \hfill \\
   - \bar \beta _3 (t)2\sum\limits_{k > 0} {g_k^2 \cos (2kd)} I_k (\Omega ,t) \hfill \\
\end{gathered}
\end{equation}

\begin{equation}\label{A10}
\begin{gathered}
  \frac{{d\bar \beta _2 }}
{{dt}} =  - \bar \beta _2 (t)\sum\limits_k {\left( {g_k^{(0)} }
\right)^2 } I_k (\Omega _0 ,t) + i\frac{{\Omega  - \Omega _0 }}
{2}\bar \beta _2 (t) \hfill \\
   - \bar \beta _1 (t)2\sum\limits_{k > 0} {g_k^{(0)} g_k^{} \cos (kd)} I_k (\Omega ,t) \hfill \\
   - \bar \beta _3 (t)2\sum\limits_{k > 0} {g_k^{(0)} g_k^{} \cos (kd)} I_k (\Omega ,t) \hfill \\
\end{gathered}
\end{equation}

\begin{equation}\label{A11}
\begin{gathered}
  \frac{{d\bar \beta _3 }}
{{dt}} =  - \bar \beta _3 (t)\sum\limits_k {g_k^2 } I_k (\Omega
,t) - i\frac{{\Omega  - \Omega _0 }}
{2}\bar \beta _3 (t) \hfill \\
   - \bar \beta _1 (t)2\sum\limits_{k > 0} {g_k^2 \cos (2kd)} I_k (\Omega ,t) \hfill \\
   - \bar \beta _2 (t)2\sum\limits_{k > 0} {g_k^{} g_k^{(0)} \cos (kd)} I_k (\Omega _0 ,t) \hfill \\
\end{gathered}
\end{equation}

The next step is to relate the coupling constants $g_k$ to the
qubit decay rate of spontaneous emission into waveguide mode. In
accordance with Fermi golden rule we define the qubit decay rates
by the following expressions:

\begin{equation}\label{A12a}
\Gamma  = 2\pi \sum\limits_k {g_k^2 \delta (\omega _k  - \Omega )}
\end{equation}

\begin{equation}\label{A12b}
\Gamma _0  = 2\pi \sum\limits_k {\left( {g_k^{(0)} } \right)^2
\delta (\omega _k  - \Omega _0 )}
\end{equation}

where

\begin{equation}\label{A13}
g_k^{}  = \sqrt {\frac{{\omega _k D_{} ^2 }} {{2\hbar \varepsilon
_0 V}}},
\end{equation}
$D$ is the matrix element of the qubit's dipole moment operator,
$V$ is the effective volume where the interaction between qubit
and electromagnetic field takes place. For 1D case, a summation
over $k$ is replaced by the integration over $\omega$  in
accordance with the prescription:

\begin{equation}\label{A14}
\sum\limits_k {}  \Rightarrow \frac{L} {{2\pi }}\int\limits_{ -
\infty }^\infty  {dk}  = \frac{L} {{2\pi }}2\int\limits_0^\infty
{d\left| k \right|}  = \frac{L} {{\pi \upsilon _g
}}\int\limits_0^\infty  {d\omega _k }
\end{equation}

where $L$ is a length of the waveguide, and we assumed a linear
dispersion law,  $\omega_k=v_gk$. The application of (\ref{A14})
to, for example, (\ref{A12a}), allows the relation between a
coupling constant $g_k$ and the decay rate $\Gamma$.

\begin{equation}\label{A15a}
g_\Omega ^{}  = \sqrt {\frac{{\Omega D_{} ^2 }} {{2\hbar
\varepsilon _0 V}}}  = \left( {\frac{{v_g \Gamma }} {{2L}}}
\right)^{1/2}
\end{equation}

Therefore, we may relate the coupling constants $ g_\Omega ^{(0)}
,g_{\Omega _2 }^{(0)} ,g_{\Omega _0 }^{}$ with their respective
decay rates.
\begin{equation}\label{A15b}
g_\Omega ^{(0)}  = \sqrt {\frac{{\Omega D_0 ^2 }} {{2\hbar
\varepsilon _0 V}}}  = \left( {\frac{\Omega } {{\Omega _0 }}}
\right)^{1/2} \left( {\frac{{v_g }} {{2L}}\Gamma _0 }
\right)^{1/2}
\end{equation}

\begin{equation}\label{A15c}
g_{\Omega _0 }^{}  = \sqrt {\frac{{\Omega _0 D_{} ^2 }} {{2\hbar
\varepsilon _0 V}}}  = \left( {\frac{{\Omega _0 }} {\Omega }}
\right)^{1/2} \left( {\frac{{v_g }} {{2L}}\Gamma } \right)^{1/2}
\end{equation}

\begin{equation}\label{A15d}
g_{\Omega _0 }^{(0)}  = \sqrt {\frac{{\Omega _0 D_0 ^2 }} {{2\hbar
\varepsilon _0 V}}}  = \left( {\frac{{v_g }} {{2L}}\Gamma _0 }
\right)^{1/2}
\end{equation}

Now we can calculate the different terms in
(\ref{A9})-(\ref{A11}). We begin with the sum in the first line in
(\ref{A9}).

\begin{equation}\label{A16}
\begin{gathered}
  \sum\limits_k {g_k^2 } I_k (\Omega ,t) = \sum\limits_k {g_k^2 } \left( {\pi \delta (\omega _k  - \Omega ) - iP.v.\left( {\frac{1}
{{\omega _k  - \Omega }}} \right)} \right) \hfill \\
   = \frac{\Gamma }
{2} - iP.v.\sum\limits_k  \left( {\frac{{g_k^2 }} {{\omega _k  -
\Omega }}} \right) \approx \frac{\Gamma }
{2} \hfill \\
\end{gathered}
\end{equation}
where $\Gamma$ is given in (\ref{A12a}).

The second term in (\ref{A16}) gives rise to the shift of the
qubit frequency. Therefore, we incorporate it in the renormalized
qubit frequency and will not write it explicitly any more. The sum
in the second line in (\ref{A9}) is calculated as follows:

\begin{equation}\label{A17}
\begin{gathered}
  2\sum\limits_{k > 0} {g_k^{} g_k^{(0)} \cos (kd)} I_k (\Omega _0 ,t) =  \hfill \\
   = \frac{L}
{{\upsilon _g }}\int\limits_0^\infty  {g_k^{(0)} g_k^{} \cos (kd)\delta (\omega _k  - \Omega _2 )d\omega _k }  \hfill \\
   - 2i P.v.\sum\limits_{k > 0}  \left( {\frac{{g_k^{} g_k^{(0)} \cos (kd)}}
{{\omega _k  - \Omega _{_0 } }}} \right) \hfill \\
   = \frac{L}
{{\upsilon _g }}g_{_{\Omega 0} }^{(0)} g_{\Omega _0 }^{} \cos (k_0
d) - i\frac{L} {{v_g \pi }}g_{\Omega _0 }^{} g_{\Omega _0 }^{(0)}
P.v.\int\limits_0^\infty  {d\omega} \frac{{\cos \left(
{\frac{\omega } {{v_g }}d} \right)}}
{{\omega  - \Omega _0 }} \hfill \\
\end{gathered}
\end{equation}

For principle value integral in (\ref{A17}) we obtain with a good
accuracy (see Appendix B):
\begin{equation}\label{A18}
P.v.\int\limits_0^\infty  {d\omega} \frac{{\cos \left(
{\frac{\omega } {{v_g }}d} \right)}} {{\omega  - \Omega }} \approx
- \pi \sin \left( {\frac{\Omega } {{v_g }}d} \right) =  - \pi \sin
\left( {k_\Omega  d} \right)
\end{equation}

Therefore, we finally obtain:
\begin{equation}\label{A19}
\begin{gathered}
  2\sum\limits_{k > 0} {g_k^{} g_k^{(0)} \cos (kd)} I_k (\Omega _0 ,t) = \frac{L}
{{\upsilon _g }}g_{_{\Omega _0 } }^{(0)} g_{\Omega _0 }^{} e^{ik_0 d}  \hfill \\
   = \frac{1}
{2}\left( {\frac{{\Omega _0 }}
{\Omega }} \right)^{1/2} \sqrt {\Gamma \Gamma _0 } e^{ik_0 d}  \hfill \\
\end{gathered}
\end{equation}

Similar calculations give for the last sum in (\ref{A9}):

\begin{equation}\label{A20}
2\sum\limits_{k > 0} {g_k^2 \cos (2kd)} I_k (\Omega ,t) =
\frac{\Gamma } {2}e^{2ikd}
\end{equation}

Collecting together (\ref{A16}), (\ref{A19}), and (\ref{A20}) we
write the final form of equation (\ref{A9}):

\begin{equation}\label{A21}
\begin{gathered}
  \frac{{d\bar \beta _2 }}
{{dt}} =  - \frac{{\Gamma _0 }} {2}\bar \beta _2 (t) +
i\frac{{\Omega  - \Omega _0 }}
{2}\bar \beta _2 (t) \hfill \\
   - \frac{1}
{2}\left( {\frac{\Omega }
{{\Omega _0 }}} \right)^{1/2} \sqrt {\Gamma \Gamma _0 } e^{ikd} \left( {\bar \beta _1 (t) + \bar \beta _3 (t)} \right) \hfill \\
\end{gathered}
\end{equation}

Similar calculations for the equations (\ref{A10}) and (\ref{A11})
yield the following result:

\begin{equation}\label{A22}
\begin{gathered}
  \frac{{d\bar \beta _2 }}
{{dt}} =  - \frac{{\Gamma _0 }} {2}\bar \beta _2 (t) +
i\frac{{\Omega  - \Omega _0 }}
{2}\bar \beta _2 (t) \hfill \\
   - \frac{1}
{2}\left( {\frac{\Omega }
{{\Omega _0 }}} \right)^{1/2} \sqrt {\Gamma \Gamma _0 } e^{ikd} \left( {\bar \beta _1 (t) + \bar \beta _3 (t)} \right) \hfill \\
\end{gathered}
\end{equation}

\begin{equation}\label{A23}
\begin{gathered}
  \frac{{d\bar \beta _3 }}
{{dt}} =  - \frac{\Gamma } {2}\bar \beta _3 (t) - i\frac{{\Omega
- \Omega _0 }} {2}\bar \beta _3 (t) - \bar \beta _1
(t)\frac{\Gamma }
{2}e^{2ikd}  \hfill \\
   - \bar \beta _2 (t)\frac{1}
{2}\left( {\frac{{\Omega _0 }}
{\Omega }} \right)^{1/2} \sqrt {\Gamma \Gamma _0 } e^{ik_0 d}  \hfill \\
\end{gathered}
\end{equation}
\\\\

\section{Proof of equation (\ref{A18})}
The integral \[ \int\limits_0^\infty  {d\omega } \frac{{\cos
\left( {\frac{\omega } {{v_g }}d} \right)}} {{\omega  - \Omega }}
= \int\limits_0^\infty  {dx} \frac{{\cos \left( {ax} \right)}} {{x
- 1}}
\],
where $a=k_\Omega d$, can be expressed in terms of sine and cosine
integrals, $ci$ and $si$ \cite{Prud}:
\begin{equation}\label{B1}
\begin{gathered}
  \int\limits_0^\infty  {dx} \frac{{\cos \left( {ax} \right)}}
{{x - 1}} =  - \cos (a)ci(a) - \sin (a)[si(a) + \pi ] \hfill \\
   =  - \cos (a)Ci(a) - \sin (a)[Si(a) + \frac{\pi }
{2}] \hfill \\
\end{gathered}
\end{equation}

where
\begin{equation}\label{B2}
Ci(a) =  - \int\limits_a^\infty  {} \frac{{\cos t}} {t}dt;\;Si(a)
= \int\limits_0^a {} \frac{{\sin t}} {t}dt
\end{equation}

The integrand in left hand side in (\ref{B1}) has a singular point
at $x=1$ which manifests itself as a singularity of $Ci(a)$ at
$a\rightarrow 0$ in right hand side in (\ref{B1}). Therefore, we
calculate integral (\ref{B1}) as Cauchy principal value integral.

\begin{equation}\label{B3}
\begin{gathered}
P.v.  \int\limits_0^\infty  {\frac{{\cos ax}} {{x - 1}}} dx\approx
P.v.\int\limits_{ - \infty }^\infty {\frac{{\cos ax}} {{x - 1}}}
dx = P.v.\int\limits_{ - \infty }^\infty  {\frac{{\cos a(t + 1)}}
{t}} dt \hfill \\
   = \cos a\;P.v.\int\limits_{ - \infty }^\infty  {\frac{{\cos at}}
{t}} dt - \sin a\;P.v.\int\limits_{ - \infty }^\infty
{\frac{{\sin at}}
{t}} dt \hfill \\
   \hfill \\
\end{gathered}
\end{equation}

\begin{equation}\label{B4}
P.v.\int\limits_{ - \infty }^\infty  {\frac{{\cos at}} {t}} dt =
\mathop {\lim }\limits_{R \to \infty } \mathop {\lim
}\limits_{\varepsilon  \to 0} \left( {\int\limits_{ - R}^{ -
\varepsilon } {} \frac{{\cos at}} {t}dt + \int\limits_\varepsilon
^R {} \frac{{\cos at}} {t}dt} \right)
\end{equation}

\begin{equation}\label{B5}
P.v.\int\limits_{ - \infty }^\infty  {\frac{{\sin at}} {t}} dt =
\mathop {\lim }\limits_{R \to \infty } \mathop {\lim
}\limits_{\varepsilon  \to 0} \left( {\int\limits_{ - R}^{ -
\varepsilon } {} \frac{{\sin at}} {t}dt + \int\limits_\varepsilon
^R {} \frac{{\sin at}} {t}dt} \right)
\end{equation}

Since
\[
\int\limits_{ - R}^{ - \varepsilon } {} \frac{{\cos a t}} {t}dt =
- \int\limits_\varepsilon ^R {} \frac{{\cos a t}} {t}dt
\]
and
\[
\int\limits_{ - R}^{ - \varepsilon } {} \frac{{\sin a t}} {t}dt =
\int\limits_\varepsilon ^R {} \frac{{\sin a t}} {t}dt
\]
we obtain

\begin{equation}\label{B6}
P.v.\int\limits_{ - \infty }^\infty  {\frac{{\cos a t}} {t}} dt =
0
\end{equation}

\begin{equation}\label{B7}
\begin{gathered}
  P.v.\int\limits_{ - \infty }^\infty  {\frac{{\sin at}}
{t}} dt = \mathop {\lim }\limits_{R \to \infty } \mathop {\lim
}\limits_{\varepsilon  \to 0} \left( {2\int\limits_\varepsilon ^R
{} \frac{{\sin at}}
{t}dt} \right) =  \hfill \\
  \mathop {\lim }\limits_{R \to \infty } \left( {2\int\limits_0^R {} \frac{{\sin at}}
{t}dt} \right) = \mathop {\lim }\limits_{R \to \infty } \left( {2Si(aR)} \right) = \pi  \hfill \\
\end{gathered}
\end{equation}

The final result in (\ref{B7}) is due to the known relation
\cite{Prud}: $ \mathop {\lim }\limits_{x \to \infty } Si(x) =
\frac{\pi } {2}$.

Therefore, we finally obtain
\begin{equation}\label{B8}
  P.v.\int\limits_0^\infty  {\frac{{\cos ax}}
{{x - 1}}} dx \approx  - \pi \sin a
\end{equation}

Below, in Fig.\ref{Fig12} we compare the $kd$- dependence of
(\ref{B1}) with that of (\ref{B8}). We see a noticeable
discrepancy for $kd<\pi/4$. For $kd>\pi/2$ two curves are almost
identical.

\begin{figure}[h]
  \includegraphics[width=8 cm]{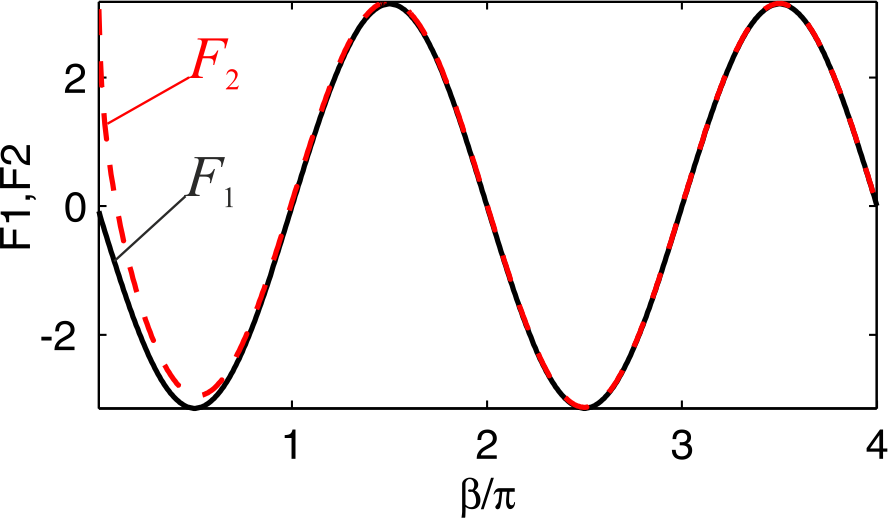}\\
  \caption{Comparison of principle value $F_1$ (right hand side of (\ref{B8})  with the
  "exact" expression $F_2$ (right hand side of (\ref{B1})).}\label{Fig12}
\end{figure}

\end{document}